\documentclass[12pt]{article}
\usepackage{stmaryrd}
\usepackage{amssymb}
\usepackage{amsfonts}
\usepackage{amsmath}
\usepackage{mathrsfs}
\usepackage{bbm}
\usepackage[nohead]{geometry}
\usepackage[singlespacing]{setspace}
\usepackage[bottom]{footmisc}
\usepackage{endnotes}
\usepackage{graphicx}
\usepackage{rotating}
\usepackage[normalem]{ulem}
\usepackage{natbib}
\usepackage{subcaption}
\usepackage{amsmath}

\usepackage{xcolor}

\usepackage{tikz}

\usepackage{hyperref} 

\newcommand{\Red}[1]{\textcolor{red}{#1}}

\newcommand{\Blue}[1]{\textcolor{blue}{#1}}

\newtheorem{theorem}{Theorem}

\newtheorem{proposition}{Proposition}

\newtheorem{lemma}{Lemma}
\newtheorem{corollary}{Corollary}

\newtheorem{definition}{Definition}
\newenvironment{proof}[1][Proof]{\noindent\textbf{#1.} }{\hfill  \rule{0.5em}{0.5em}}

\newcommand{\E}{\mathbb{E}}

\def\P{\mathbb{P}}

\newcommand{\Prob}{\mathbb{P}}

\long\def\ignore#1{}

\def\l{\ell}

\def\?{\Red{?????????}}
\def\1{\textbf{1}}

\newcommand{\headings}[1]{ \noindent\textbf {#1}  }

\newcommand{\heading}[1]{\noindent{\bf }}

\makeatletter
\def\@biblabel#1{\hspace*{-\labelsep}}
\makeatother

\begin{document}

\title{ The Value of Information in Stopping Problems   }

\author{Ehud Lehrer\footnote{School of Mathematical Sciences, Tel Aviv University, Tel Aviv 69978, Israel, Email: lehrer@post.tau.ac.il; Lehrer acknowledges the support of grants ISF 591/21 and DFG KA 5609/1-1.}\ \ and  Tao Wang\footnote{Institute for Social and Economic Research, Nanjing Audit University, Nanjing, China. Email: tao.wang.nau@hotmail.com;
        Wang acknowledges the support of NSFC Grant \#11761141007.} \\}

\date{\today}               

\maketitle

\begin{abstract}
\footnotesize{We consider stopping problems in which a decision maker (DM) faces an unknown state of nature
and decides sequentially whether to  stop and take an irreversible action; pay a fee and obtain additional information; or wait without acquiring information.
We discuss the value and quality of information. The former is the maximal discounted expected revenue the DM can generate.
We show that among all history-dependent fee schemes, the upfront scheme (as opposed, for instance, to pay-for-use) is optimal:
it generates the highest possible value of information.
The effects on the optimal strategy of obtaining information from a more accurate source and of having a higher discount factor are distinct, as far as
expected stopping time and its distribution are concerned.
However, these factors have a similar effect in that they both enlarge the set of cases in which the optimal strategy prescribes waiting. }

\bigskip


\end{abstract}

\thispagestyle{empty}

\sloppy

\newpage
\setcounter{page}{1}


\section{Introduction}

In many real-world situations, a decision maker  (henceforth, DM) is faced with dynamic decision problems under a payoff-relevant state of nature,
which is initially unknown.
In any period, based on the information about the state of nature she has,
the DM decides  whether to take an immediate irreversible action (e.g., make an investment decision or forsake the decision problem and seek an outside option)
or to defer the action choice to the future after getting more information. Since an information provider may charge fees for the services,
obtaining additional information typically comes with a cost.
This cost may depend on the exact information provided.
The DM has to balance the expected gain from additional information and its expected costs.


The first objective of this paper is to investigate the optimal history-dependent fee scheme and the value of given information structure.
For this goal we adopt the perspective of the information provider.
Assume that information (noisy signals) is generated sequentially and independently (conditional on the state of nature)
from a known information structure and the information provider is
non-strategic regarding what information to provide, but is strategic in setting fees.
In practice, there are several commonly used fee schemes. For instance, the information provider can charge a fixed amount each period,
gradually increase/decrease the fee over time,
or simply impose an upfront fee (such as subscription fees for magazines, or private databases).
More generally, the information provider can condition the fee to be paid in a certain period on the past history of signals.
This is more general than the belief-based fee schemes, since given a prior belief,
a history of signals uniquely determines a posterior belief.

Note that each history-dependent fee scheme induces a discounted expected total payment that could be collected from the DM.
We define the value of an information structure as the highest achievable discounted total expected payment.
We characterize the history-dependent fee scheme that will maximize the expected revenue of the information provider,
or equivalently, the one that achieves the value of the information structure.
\ignore{Essentially, we have an interaction between the information provider and the DM:
the information provider moves first by committing to a fee scheme.
Then, upon observing the fee scheme, the DM maximizes her expected payoff by choosing whether or not to use the services of the
information provider and by adopting an optimal stopping/waiting strategy.
\Blue{Thus, each fee scheme induces an expected total payment that could be extracted from the DM,
 and we define the ``value" of a given information structure as the maximal expected total payment.}
 }

We show that the widely used upfront fee scheme is optimal. Under this scheme,
the DM pays a lump sum at the beginning and pays no flow costs during the course of the decision making process.
The upfront fee, which equals the value of the information structure,
makes the DM indifferent between making decisions without getting any additional information and
paying the fee to get free information for the rest of the decision process.

To elucidate the cause of this result, consider a fee scheme with positive flow costs.
When the DM decides whether to pay and acquire information in a given period,
she has to weigh the expected benefits of additional information
(which will better acquaint her with the prevailing state thus allowing for more accurate decisions) and the entire future costs that she expects to incur.
It is possible that a relatively large future flow costs would deter the DM from acquiring information in earlier stages and force her to stop prematurely.
This  effect might reduce the DM's discounted expected payoff,
because, in the absence of an accurate knowledge of the prevailing state, her action may be sub-optimal.
In contrast, the upfront fee scheme allows the DM to obtain more information before stopping,
hence generating the maximal discounted expected gross payoff for the DM.
The information provider can reap the benefit by adjusting the upfront fee.

It is worth noting that the optimality of the upfront fee holds when the information provider's discount factor is no greater than
the DM's discount factor. However, when the information provider adopts a strictly greater discount factor,
the upfront fee scheme is no longer optimal. In this case, both players benefit from trading payoffs over time,
and the optimal fee scheme involves charging a lump sum and sufficiently delaying the payment time.

The second objective is to investigate how the information quality affects the DM's optimal strategies.
When faced with a dynamic decision problem, different DMs exhibit significantly different behaviors
even when they share very similar utility functions and payoffs, and hence similar risk attitudes.
Some are more decisive and  tend to make decisions quickly once their beliefs reach some favorable level,
while others are more patient, and prefer to defer their decisions.
The classical economic explanation is that different DMs have different time preferences.
The more patient individuals have higher discount factors, which implies that the cost of waiting is lower.
We examine the effect of information quality on DM's waiting behavior and highlight the similarities and distinctions with the effect of a greater discount factor.

To this end, consider two information structures $S$ and $T$, where $S$ is more informative than $T$ in the sense of Blackwell.
We show that in every period, waiting is optimal for the DM under a larger set of beliefs when she receives information from $S$.
This effect is shared also by having a greater discount factor, namely being more patient.
However, in terms of the expected stopping time and its distribution, better information has effects that are distinct from those of a greater discount factor.
A greater discount factor induces a first-order stochastically dominating stopping time distribution,
hence it implies a longer expected stopping time. In contrast, the effect of better information is ambiguous.
On the one hand, the larger waiting sets under a Blackwell-dominating information structure $S$ tends to prolong the waiting process.
On the other hand, signals generated from $S$, being more informative, lead to a faster belief updating.
The latter effect tends to be more pronounced when the belief is bounded further away from the optimal stopping sets.
We demonstrate that, in general, a Blackwell-dominating information structure yields neither a first-order, nor a second-order
stochastically dominating stopping time distribution.

Our paper is related to a large literature on learning and sequential decision making under payoff-relevant uncertainty
initiated by \citet{Wald1945}, \citet{Wald1947}. 
Early works focus on the ``option value" generated by maintaining the choice to ``wait and see" open in various of economic applications
(for example, \citet{Arrow1974}, \citet{Cukierman1980}, \citet{Pindyck1991}, \citet{Demers1991}, \citet{Chetty2007} among many others).
In these papers, the information that arrives sequentially is generated exogenously.
This strand of literature is extended in several directions.
\citet{Moscarini2001} endogenize the choice of signal precision.
\citet{Moscarini2002} obtain a complete order of the experiments as the sample size is sufficiently large.
\citet{Che2019}, \citet{Liang2019}  and  \citet{Mayskaya2019} consider the sequential choice of information sources.
\citet{Morris2019} establish a one-to-one correspondence between the ex-ante cost function and the dynamic flow cost (both are functions of beliefs)
in the sequential sampling problem. They also show that the former equals the expected change of the log likelihood ratio.
In all listed papers, the waiting costs or information acquisition costs are given exogenously.
In our paper, by contrast, we treat such costs as endogenous and characterize the optimal history-dependent cost structure that extracts the maximal
expected surplus from a DM.

Our paper is also related to the literature on the value and cost of information.
\citet{Gilboa1991} identify necessary and sufficient conditions for a function over partitions of the state space to be a value of information function.
The analysis is extended further to stochastic information structures in \citet{Azrieli2008}.
\citet{Cabrales2013} identify a wide class of utility functions and investment problems where information structures can be ordered \emph{completely}
by the decrease in entropy of the DM’s beliefs.
\citet{DeLara2017} investigate the duality between decisions and preference on the one hand, and the value function on the other.
More recently, \citet{Frankel2019} develop axiomatic characterizations for functions that are valid (ex post) measures of information and uncertainty.
These authors provide conditions under which the expected reduction in uncertainty equals the expected amount of information generated.
In addition, they examine sequential information provision. The information provider is endowed with a given information structure in each period and will
eventually reveal all information to the DM, but the information provider can decide whether to delay the arrival of information or hide information.
Their focus is to characterize the history-dependent information cost functions
under which the information provider has no incentive to hide or delay information.
But unlike our paper, the authors do not consider the DM's stopping decisions.
\citet{Bloedel2020} consider DMs who choose experiments sequentially to minimize a direct cost function under the constraint that the
sequentially acquired information is at least as informative as a target level.
They adopt an axiomatic approach and characterize the cost functions that are sequentially learning-proof.
By contrast, our paper focuses on the optimal cost function that maximizes DMs' willingness to pay.

In addition, there is a related strand of literature which focuses on the optimal way to price and sell information.
For instance, in \citet{Bergemann2015}, a monopolistic data provider sells individual-level match value data (cookies) to advertisers.
The authors study the optimal linear pricing of cookies.
In comparison, we consider a much larger set of fee schemes, including history-dependent ones.
\citet{babaioff2012optimal}, \citet{Bergemann2018} study optimal ways to sell information when there are either incomplete information on the
side of the information buyer or the information seller. Specifically, in \citet{Bergemann2018},
a data buyer has private information about an unknown state of nature,
an information seller maximizes the expected revenue by designing a menu of information structures
(which may include the fully informative one) and setting a price for each of them (i.e., the price does not depend on signal realization).
They show that, in general, the optimal menu contains both partially distorted information structures and the fully informative one.
In our paper, unlike \citet{Bergemann2018}, the information provider is endowed with a given information structure and we consider fee schemes that
depends on past histories of signals. \citet{babaioff2012optimal} studies the optimal selling mechanism in an environment in which
both information seller and information buyer have private information (or types), the joint distribution of which is known.
The buyer's payoff depends on both players' types as well as the buyer's own action.
In our approach, there are no private types, and the DM's payoff depends on an unknown state.
Besides, the DM chooses when to stop the information acquisition process, rather than sending messages of private type, as in \citet{babaioff2012optimal}.

\ignore{Besides, there is a related strand of literature that studies the optimal way to price and sell information.
For instance, \citet{Bergemann2015} focus on selling and purchasing individual-level match value data in an online advertising environment;
\citet{babaioff2012optimal}, \citet{Bergemann2018} consider situations in which either the information buyer or the information seller have private information
and analyze the optimal way to sell information.
These works are conducted in a static environment.  }

The paper is organized as follows. Section \ref{Section_Model} sets up the model and presents a few preliminary results.
Section \ref{Section: Optimal Fee Scheme} characterizes the optimal history-dependent fee scheme and derives the value of a given information structure.
In Section \ref{Section: Comparing Info Structures}, we investigate the effects of information quality on the optimal strategies and the stopping time.  
Section \ref{Section_conclusion} concludes. Other omitted proofs can be found in the appendix.

\section{The Decision Problem}\label{Section_Model}

A DM faces a stopping problem with decision horizon $N$, $N\in\mathbb{N}\cup\{+\infty\}$,
and a time-invariant payoff-relevant state of nature $\theta$,
$\theta\in\Theta=\{\theta_1,\theta_2,...,\theta_m\}$, $m\geq 2$. Denote a generic period by $n$, $n=0,1,...,N$, and let
$\mu_0\in\Delta\Theta$ be the prior belief, where $\Delta\Theta$ is the space of all probability distributions over $\Theta$.
The DM decides sequentially among the following two options:
(i) stop immediately and take an action to collect the expected stopping payoff;
(ii) wait and acquire information about $\theta$ from a known information source (henceforth referred to as \emph{information acquisition}).
Assume the stopping decisions are irrevocable. 

\ignore{
The option to wait without acquiring information is relevant for two reasons. First, with a transition matrix $\tau$, the belief may evolve gradually to
a favorable one with sufficiently high probability, even without additional information.\footnote{Consider the mapping
$\Gamma:\Delta\Theta\rightarrow\Delta\Theta$ defined by $\Gamma(\mu)=\mu\tau$.
This mapping captures the evolution of belief based solely on the transition matrix $\tau$.
If all elements of $\tau$ are strictly less than $1$, then one can show that $\Gamma$ is a contraction mapping.
Hence, it has a unique fixed point $\mu^*$, to which the belief process (without addition information) will converge.
Sometimes convergence can be achieved immediately.
For instance, if $\tau=\big(\begin{smallmatrix}  0.8 & 0.2 \\ 0.8 & 0.2 \end{smallmatrix}\big)$, then regardless of the prior, the belief next period
will be $(0.8,0.2)$. }
Second, acquiring extra information may be costly (see the discussion below).
}

If the DM chooses to stop, her stopping payoff will be $u(a,\theta)$, where $a$ is an action that belongs to a compact action set $A$.
 In particular, $A$ may contain the action to forsake the decision problem and get the payoff of an outside option, which is normalized to $0$.
Assume the payoff function $u(a,\theta)$ is bounded and continuous in $a$, for every $\theta$.
If the DM waits beyond the termination period, we assume that she receives the payoff of the outside option.

If in period $n$, $n=0,...,N-1$, the DM chooses to acquire information for at least one more period,
she would receive a noisy signal $s$ about the prevailing state at period $n+1$ generated by a known stochastic information structure $S$
characterized by $f_{\theta}(s)$, $\theta\in\Theta$.
Based on the noisy signal $s$ obtained, the DM updates her belief,
and decides again in period $n+1$ whether to stop, wait, or acquire information.
Assume the signals are independent conditional on the state.
This sequential decision  process continues until either the DM stops or waits beyond the termination period $N$.
Assume the DM discounts future payoffs by a fixed discount factor $\delta,\ \delta\in(0,1]$.

\bigskip

\headings{Cost of Information.}
Information is usually obtained with a cost. The cost to get a signal could depend on the history of past signals obtained by the DM.
Let $h_n$ be a history of signals of length $n$ and let $H_n$ be the collection of all length-$n$ histories.
Denote $\mathcal{H}_N:=\left(\bigcup_{n=1}^N H_n\right)\bigcup \varnothing$ as the set of all histories,
where $\varnothing$ represents the null history.

For a given prior belief $\mu_0$,
the cost of acquiring information from the information structure is represented by a \emph{fee scheme}
$c: \mathcal{H}_N\rightarrow\mathbb{R}_+$ that specifies a fee to be paid for each possible history.\footnote{If the signal space
$\mathcal{S}$ has $K-1$ elements, then a fee function $c$ can be represented by a point in $\mathbb{R}_+^Z$, where $Z=1+K+\cdots+K^N=\frac{1}{K-1}(K^{N+1}-1)$.}
Since for a fixed prior belief a history uniquely determines the posterior, history-dependent fee functions are more general than belief-based fee schemes.
Notice that $c(h_n)$ is not the cumulative expenditure, but rather the amount to be paid in period $n$ given a history $h_n$.
For instance, in one special case, $c(\varnothing)>0$, while $c(h_n)=0$ for all $h_n\in\mathcal{H}_N\backslash \varnothing$.
This corresponds to the \emph{upfront fee scheme}, in which the DM only needs to pay a lump sum $c(\varnothing)$ in period 0 to access all future information.
In another special case, $c(\varnothing)=0$, while $c(h_n)=c>0$ for all $h_n\in\mathcal{H}_N\backslash \varnothing$.
This represents the flat-rate fee scheme.

\bigskip

\headings{Value Functions and Stopping Strategies.}
Fix a prior belief $\mu_0$, an information structure $S$, and a fee function $c(\cdot)$.
If the DM holds belief $\mu$, $\mu=(\mu_\theta)_{\theta\in\Theta}$,\footnote{Here $\mu$ is treated as a row vector.} in a certain period,
then her maximal expected stopping payoff is (when the action set $A$ includes the action to forsake the problem and get the outside option,
$\pi(\mu)$ is non-negative)
\begin{equation*}
\pi(\mu):=\max_{a\in A} \sum_{\theta\in\Theta}u(a,\theta)\mu_\theta;
\end{equation*}
and if she acquires information, she would receive the discounted expected continuation value minus the discounted expected fee to be paid next period.
Formally, let $V^N_n(\mu;h_n)$ be the period-$n$ value function given a history $h_n$.
The dependence on $h_n$ is due to the fee function.
The discounted continuation payoff, if the DM acquires information, is
$\delta \left[ \E \left( V^N_{n+1}(\mu(s);h_{n+1}) | \mu; h_n \right) - \E(c(h_{n+1})|\mu; h_n) \right] $,
where expectations are with respect to the signal $s$ to be obtained in period $n+1$,
and $\mu(s)$ is the posterior belief after getting signal $s$.
Hence, the period-$n$ value function $V^N_n(\mu;h_n)$ can be written as
\begin{equation}\label{Value_Func_V_n}
V_n^N(\mu;h_n;c) =  \max \Big\{ \pi(\mu), \delta \left[ \E \left( V^N_{n+1}(\mu(s);h_{n+1}) | \mu; h_n \right) - \E(c(h_{n+1})|\mu; h_n) \right] \Big\}.
\end{equation}

Conditional on each history $h_n$, the optimal stopping and information acquisition strategies in period $n$,
$1\leq n\leq N-1$, are determined by the comparison of the two terms in the expression of the value function.
We therefore define the \emph{stopping set}, and \emph{information acquisition set} following a history $h_n$ as
$ E^N_n(h_n):=\left\{ \mu\in\Delta\Theta |~  V^N_n(\mu;h_n)=\pi(\mu)  \right\}$,
and $A^N_n(h_n):=\Delta\Theta\backslash E^N_n(h_n)$, respectively.

When the flow cost is a constant, i.e., $c(h_n)=c$ for all $h_n$,
the sequences of value functions $\{V_n^N\}_{n=0}^N$ and stopping sets $\{E^N_n\}_{n=0}^N$ have the following property.
\begin{proposition}\label{Prop_decreasing stopping sets}
Suppose the flow cost is a constant $c$. Then for any belief $\mu\in\Delta\Theta$, $V^N_n(\mu)\geq V^N_{n+1}(\mu)$ and $E^N_n\subseteq E^N_{n+1}$.
\end{proposition}
The proposition says that other things being equal, a DM who faces a longer decision horizon receives a greater discounted expected payoff.
Consequently, as time approaches the termination date, the DM adopts a less stringent stopping criterion.


\section{The Optimal Fee Scheme}\label{Section: Optimal Fee Scheme}
The purpose of this section is to study the optimal history-dependent fee scheme from the perspective of the information provider and derive the value of a given
information structure in the stopping problem described in Section \ref{Section_Model}.
In the literature that studies sequential decision problems, the cost to get information each time is given exogenously.
Here, the cost is endogenous.
We know that each history-dependent fee scheme induces a sequence of optimal stopping and information acquisition strategies.
Using the optimal strategies, one can calculate the discounted expected amount of money to be collected from the DM during the decision making process.
We define this maximal total discounted expected payment as the ``value" of a given information structure. 
Essentially, we have an interaction between the information provider (``he") and the DM (``she"):
in period 0, the information provider chooses a history-dependent fee scheme $c$ and commits to it,
and then the DM decides whether to take it or leave it. Assume the information provider is non-strategic regarding what information to provide.
We consider the case in which the DM and the information provider adopt the same discount factor $\delta$.
In Subsection \ref{Section_Optimal Fees_Extensions}, we discuss what happens when their discount factors are different.

If the DM declines the fee scheme proposed by the information provider
(for instance, due to a sufficiently pessimistic/optimistic prior belief, forbiddingly high fees),
then she just relies on her prior belief $\mu_0$ and chooses an action in $A$ to get expected payoff $\pi(\mu_0)$.\footnote{Since the state of nature is time-invariant,
if the DM chooses not to get information, her belief would remain the same.
Due to discounting, the DM will not delay her action. }
But if the DM accepts the fee scheme, then she pays an upfront fee $c(\varnothing)$, if any, and waits at least one period.
Therefore, the discounted expected continuation payoff from acquiring information is
$\delta \left[ \E \left( V^N_{1}(\mu_0(s);h_{1};c) | \mu_0; \varnothing \right) - \E(c(h_{1})|\mu_0; \varnothing)\right] - c(\varnothing)$.
Hence the period-$0$ value function can be written as:
\begin{equation}\label{Value_Func_V_0}
V_0^N(\mu_0;\varnothing;c) =  \max\Big\{ \pi(\mu_0), \delta \left[ \E \left( V^N_{1}(\mu_0(s);h_{1};c) | \mu_0; \varnothing \right) - \E(c(h_{1})|\mu_0; \varnothing) \right] - c(\varnothing) \Big\}.
\end{equation}
The period-1 value function $V^N_1(\cdot)$ satisfies Eq. (\ref{Value_Func_V_n}).

The DM will accept a fee scheme if the value function $V_0^N(\mu_0;\varnothing;c)$
is greater than her expected stopping payoff when she declines the fee scheme.
Formally, fix a stopping problem with a prior belief $\mu_0$, define
\begin{equation}\label{Set C}
C(\mu_0):=\left\{c:\mathcal{H}_N\rightarrow\mathbb{R}_+ \ \big{|} \ V^N_0(\mu_0;\varnothing; c) > \pi(\mu_0)  \right\}.
\end{equation}
The set $C(\mu_0)$ consists of all those fee schemes under which the DM with prior belief $\mu_0$ strictly prefers to accept the fee scheme and
acquire information.
The set $C(\mu_0)$ may be empty, for instance, when the prior belief is either highly pessimistic/optimistic.
In what follows, we focus on the non-trivial case in which $C(\mu_0)$ is not empty.
Note that $C(\mu_0)$ is typically unbounded.
For instance, consider a fee scheme that charges a very small amount in the first few periods and a huge amount in later periods.

Our objective is to characterize the fee scheme in $C(\mu_0)$ that maximizes the information provider's discounted expected payoff.
Fix a fee scheme $c$, $c\in C(\mu_0)$. Note that a fee is generated only when the DM chooses to acquire information.
Therefore, by using the optimal information acquisition sets $\{A^N_n(h_n;c)\}_{h_n\in H_n,0\leq n\leq N}$,
one can calculate the discounted expected total fee $\rho(\mu_0;c)$ that the DM expects to incur \emph{ex ante}
in order to get information from the information structure $S$.
Formally, for a history $h_n$, denote its predecessor with length $m$ by $h_{n,m}$, $0\leq m<n$.
The discounted expected total fee being incurred up to time $N$ is
\begin{equation}\label{Total expected cos under c}
\rho(\mu_0;c) := \sum_{n=0}^{N} \sum_{h_n\in H_n} \delta^n  c(h_n) \Prob\left(h_n|\mu_0\right) \mathbbm{1}_{\left\{\mu_0(h_{n,m})\notin E^N_m(h_{n,m};c),\ \mu_0(h_n)\in A^N_n(h_n;c) \right\}}.
\end{equation}
The term $\Prob\left(h_n|\mu_0\right)$ is the probability of observing a history $h_n$, given a prior belief $\mu_0$.
Notice that a fee $c(h_n)$ is paid at period $n$ when the DM chooses to acquire information at period $n-1$, after observing the predecessor $h_{n,n-1}$.

\begin{definition}\label{Def_Value of Info}
Given a decision problem and a prior belief $\mu_0$, the \emph{value} of an information structure $S$ to the \emph{DM} is
\begin{equation*}\label{Value of Info}
\rho^*(\mu_0) := \sup_{c\in C(\mu_0)} \rho(\mu_0;c).
\end{equation*}
\end{definition}

For fee functions outside the set $C(\mu_0)$, since the DM can always choose not to get any information,
she will base her decision solely on the prior belief, so we have $\rho(\mu_0;c)=0$.

To proceed, we first state the following lemma that decomposes the value function. 
We use the notation ``$\pmb{0}$" to denote the zero fee scheme, i.e., the one that charges zero fee following every history. 

\begin{lemma}\label{Lemma_Decompose value function}
The period-$0$ value function under a given fee function $c\in C(\mu_0)$ can be decomposed as
\begin{equation}\label{Decompose value function}
V^N_0(\mu_0;\varnothing;c) = \hat{V}^N_0(\mu_0;\varnothing;\pmb{0}) - \rho(\mu_0;c),
\end{equation}
where $\hat{V}^N_0(\mu_0;\varnothing;\pmb{0})$ is the discounted expected payoff when the \emph{DM} faces the fee function $\pmb{0}$, but
follows the optimal strategy corresponding to the fee function $c$.\ignore{ \footnote{The difference between $\hat{V}^N_0(\mu_0;\varnothing;\pmb{0})$ and
$V^N_0(\mu_0;\varnothing;\pmb{0})$ is as follows: $V^N_0(\mu_0;\varnothing;\pmb{0})$ is the expected payoff when $c=\pmb{0}$ and the DM follows
the optimal strategy corresponding to $c=\pmb{0}$; in contrast, $\hat{V}^N_0(\mu_0;\varnothing;\pmb{0})$ is the expected payoff when $c=\pmb{0}$,
but the DM follows the optimal strategy corresponding to a fee function $c$. } }
\end{lemma}

Lemma \ref{Lemma_Decompose value function} says that the value the DM can expect at the beginning of the decision problem
when facing a fee function $c$ can be expressed as the expected \emph{gross} payoff minus the discounted expected total costs.
The proof is relegated to the Appendix. 

The information provider faces the following trade-off. On the one hand, charging a high fee following each history
has the benefit of increasing his own immediate profit, but it might deter the DM from acquiring more information.
On the other hand, charging a low fee following each history has the advantage of encouraging the DM to acquire information more frequently.
The following result  clarifies the dilemma  and  characterizes the structure of the optimal fee scheme.

\begin{theorem}\label{Theorem_Compare Value of Info}
Given any information structure $S$ and a prior belief $\mu_0\in \Delta\Theta$,
we have $\rho^*(\mu_0) = V^N_0(\mu_0;\varnothing;\pmb{0}) -\pi(\mu_0)$.
Thus, the value can be achieved via the fee scheme that only charges an upfront fee equal to $V^N_0(\mu_0;\varnothing;\pmb{0}) -\pi(\mu_0)$.
\end{theorem}
\begin{proof}
First observe that $\hat{V}^N_0(\mu_0;\varnothing;\pmb{0})\leq V^N_0(\mu_0;\varnothing;\pmb{0})$.
Indeed, by definition, $V^N_0(\mu_0;\varnothing;\pmb{0})$ is the \emph{maximal} discounted payoff
that the DM can expect under fee function $\pmb{0}$,
whereas $\hat{V}^N_0(\mu_0;\varnothing;\pmb{0})$ is the expected payoff when the DM faces the fee function $\pmb{0}$, but
follows the optimal strategy corresponding to the fee function $c$.

Now let $\varphi(\mu_0):=V^N_0(\mu_0;\varnothing;\pmb{0}) -\pi(\mu_0)$, which is the maximal upfront fee that can be imposed.
Take any $c\in C(\mu_0)$. We show that $\varphi(\mu_0)\geq\rho(\mu_0;c)$.
Using Lemma \ref{Lemma_Decompose value function} and the definition of the set $C(\mu_0)$
(see Eq.\ (\ref{Set C})), we have
\begin{equation}\label{equation_proof theorem_value of info_1}
\pi(\mu_0) < V^N_0(\mu_0;\varnothing;c) = \hat{V}^N_0(\mu_0;\varnothing;\pmb{0}) - \rho(\mu_0;c).
\end{equation}
Using the definition of $\varphi(\mu_0)$ and rearranging the terms, we obtain
\begin{equation}\label{equation_proof theorem_value of info_2}
\pi(\mu_0) = V^N_0(\mu_0;\varnothing;\pmb{0}) - \varphi(\mu_0).
\end{equation}
Eqs.\ (\ref{equation_proof theorem_value of info_1}) and  (\ref{equation_proof theorem_value of info_2}) yield
\begin{equation*}
\hat{V}^N_0(\mu_0;\varnothing;\pmb{0}) - \rho(\mu_0;c) > V^N_0(\mu_0;\varnothing;\pmb{0}) - \varphi(\mu_0),
\end{equation*}
which in conjunction with the fact that $\hat{V}^N_0(\mu_0;\varnothing;\pmb{0}) \leq V^N_0(\mu_0;\varnothing;\pmb{0})$  implies that
\begin{equation*}
\varphi(\mu_0) - \rho(\mu_0;c) >  V^N_0(\mu_0;\varnothing;\pmb{0}) - \hat{V}^N_0(\mu_0;\varnothing;\pmb{0}) \geq 0.
\end{equation*}
Since this inequality holds for every $c\in C(\mu_0)$,
we conclude that $\varphi(\mu_0) \geq \sup_{c\in C(\mu_0)} \rho(\mu_0;c) = \rho^*(\mu_0)$.
This completes the proof.
\end{proof}

\bigskip

\ignore{ The reason behind this theorem is as follows.
When the DM faces a generic
fee function $c$, then according to Lemma \ref{Lemma_Decompose value function},
her expected gross payoff equals the expected payoff when the DM pays no fee, but follows the optimal strategy corresponding to the fee function $c$.
Typically, this optimal strategy is different from the optimal strategy under fee function $\pmb{0}$.
As a result, the expected gross payoff under a generic fee scheme $c$ is lower than that under $\pmb{0}$,
which leads to a lower willingness to pay.
Put it differently, fee functions other than the upfront fee scheme will create distortions in the DM's optimal waiting, entry and exit strategies,
which reduces the expected gross payoff and leads to inefficient outcomes.
Our result suggests that in practice, the widely used ``subscription fee" payment method could be efficient in a wide range of dynamic decision problems.}

To understand the reason behind this result, consider a fee function that charges a non-negative amount following every history.
Each time the DM decides whether to stop or acquire information, she balances the expected benefit from additional information and the expected cost.
Additional information is always beneficial in this environment, since on expectation belief is updated to the right direction, which helps the DM to make more
accurate decisions. However, the presence of positive flow fee deters the DM from getting more information and leads to earlier stopping.
Hence, upon stopping, the DM may not know the prevailing state sufficiently well as she would were she to wait longer and acquire more information.
Consequently, on average, the action taken upon stopping tends to be further away from the optimal action under the true state.
This is inefficient and it leads to a lower discounted expected gross payoff for the DM.
In comparison, under the upfront fee scheme, the DM faces no cost consideration during the course of the decision making process.
Hence the maximal discounted expected gross payoff can be achieved.
The information provider then reaps the benefit of greater discounted expected gross payoff by charging a higher upfront fee.

From the discussion above, it is also clear that the upfront fee $\varphi(\mu_0)$ not only equals the value of the information structure,
but is also the maximal welfare of the two players. Sometimes, the information provider may charge an upfront fee less than $\varphi(\mu_0)$,
for instance, due to the extra bargaining power of the DM. But as long as the upfront fee scheme is used, there is no welfare loss.
For a given problem, the upfront fee may not be the unique optimal one. But generically, it is the only fee scheme that is optimal for every problem.

It is worth noting that the optimality of the upfront fee scheme does not depend on the information structure under consideration.
Thus, regardless of the information structure, the value can be achieved via this scheme.
Also, Theorem \ref{Theorem_Compare Value of Info} is not driven by the assumption that future payoffs are discounted.
In fact, even when payoffs are undiscounted, the result still holds.

At first glance, the optimality of the upfront fee scheme in our paper appears to be similar to the following well-known result on the
principal-agent problem (both when the agent's action is observable and unobservable).
When the agent is risk neutral,  the principal (information provider) can achieve the first best by ``selling the project to the agent"
for a fixed price (the upfront fee) and make the agent (DM) the residual claimant. 
In the principal-agent problem, the agent would incur greater disutility when taking an action that benefits the principal more.
The driving force of the result is \emph{risk sharing}:
The optimal compensation scheme should  balance between \emph{risk sharing} and \emph{incentives}. While
the former ensures that the agent's wage payment does not depend heavily on the outcome, the latter awards/punishes the agent for good/bad outcomes.
When the agent is risk neutral, the risk sharing concerns disappear and the agent does not mind bearing all the output risk,
hence selling the project upfront to the agent is efficient.  The rationale is different from that of our paper.

In our model, the DM benefits from knowing the state better, and
the optimality of upfront stems from distortion considerations related to the stopping strategy.
In a wide range of stopping problems, the upfront fee is the only form of fee scheme that creates no distortion in the DM's
optimal stopping strategies. It induces the DM to wait longer and make more precise decisions.
An important implication of these different driving forces is the following.
In principal-agent problems with observable actions\footnote{Our framework is more closely related to the
principal-agent problem with observable actions, since the DM's decisions can be observed by the information provider (the principal).},
any compensation scheme that shares the same expectation as the optimal upfront fee is optimal. In contrast,
 in our paper, the upfront fee scheme is generically the unique optimal one.

\subsection{Extensions}\label{Section_Optimal Fees_Extensions}

\headings{Different Discount Factors.}
So far we assumed that the information provider and the DM share a common discount factor.
One may wonder what happens when they discount their future payoffs differently,
in particular, whether Theorem \ref{Theorem_Compare Value of Info} still holds.
Let $\delta_{\text{DM}}$ and $\delta_{\text{IP}}$ be the discount factors of the DM and the information provider (IP), respectively.

Suppose that $\delta_{\text{DM}}>\delta_{\text{IP}}$.
Note that the maximal upfront fee $\varphi(\mu_0)=V^N_0(\mu_0;\varnothing;\pmb{0})-\pi(\mu_0)$ depends only on the DM's discount factor.
So regardless of the information provider's discount factor, $\varphi(\mu_0)$ is the same.
Now fix a fee function $c$,
and let $\rho_{\text{DM}}$, $\rho_{\text{IP}}$ be the discounted expected total fees calculated using discount factors $\delta_{\text{DM}}$,
$\delta_{\text{IP}}$, respectively, as in Eq. (\ref{Total expected cos under c}).
Since the DM's optimal strategies are unaffected by the information provider's discount factor,
$\delta_{\text{DM}}>\delta_{\text{IP}}$ implies that $\rho_{\text{DM}}\geq \rho_{\text{IP}}$ for every fee function.
We therefore conclude that when $\delta_{\text{DM}}>\delta_{\text{IP}}$, the upfront fee $\varphi(\mu_0)$ is still optimal.

When $\delta_{\text{DM}}<\delta_{\text{IP}}$, imposing an upfront fee $\varphi(\mu_0)$ is no longer optimal.
To see this, consider a problem with a fixed decision horizon $N$. 
Assume that $\varphi(\mu_0)>0$.
Suppose the information provider only charges a fixed amount $\phi$ in period $K$ if the DM chooses to acquire information,
regardless of the realized history of signals (alternatively, one can interpret $\phi$ as an ``upfront fee", but paid with a $K$-period delay).
Choose $K$ and $\phi$ such that $ \delta_{\text{DM}}^K \phi \leq \varphi(\mu_0)$ and that $\delta_{\text{IP}}^K \phi > \varphi(\mu_0)$,
or equivalently, $\delta_{\text{DM}} \leq \left(\frac{\varphi(\mu_0)}{\phi}\right)^{\frac{1}{K}} < \delta_{\text{IP}}$.
The first inequality ensures that it is optimal for the DM to accept the fee scheme and acquire information,
the second one guarantees a higher discounted expected payoff for the information provider than the upfront fee $\varphi(\mu_0)$.
Clearly, for given values of $\delta_{\text{DM}},\ \delta_{\text{IP}}$ and $\varphi(\mu_0)$,
one can always find $\phi$ and $K$ such that the inequalities are satisfied.

Since the upfront fee is no longer optimal, the natural questions are what is the optimal fee scheme when $\delta_{\text{DM}}<\delta_{\text{IP}}$,
and what is the maximal discounted expected total fee that can be collected?
To proceed, note that the optimal payment scheme when $\delta_{\text{DM}}<\delta_{\text{IP}}$ must take the form of a fixed payment with a payment delay.
This is because, as in the case with a common discount factor, any other history-dependent payment scheme (even with payment delay) would lead to distortions
in the DM's optimal waiting strategies, yielding a lower value function.

Now let us consider the optimal fixed payment with delay and the maximal discounted expected total fee. The information provider's problem becomes
\begin{eqnarray*}
  &           & \max_{K, \phi}\ \  \delta_{\text{IP}}^K \phi   \\
  &\text{s.t.}& \delta_{\text{DM}}^K \phi\leq \varphi(\mu_0),  \\
  &           & K\in \mathbb{N}_+,\ \ \phi\geq 0.
\end{eqnarray*}
For the optimal solution, the constraint must bind, otherwise, one can increase the objective function by increasing $\phi$.
The optimization problem is then equivalent to
$\max_{K\in \mathbb{N}_+} \left(\frac{\delta_{\text{IP}}}{\delta_{\text{DM}}}\right)^K\varphi(\mu_0)$.
Since $\delta_{\text{IP}}>\delta_{\text{DM}}$, the solution is unbounded, implying that the information provider can achieve an arbitrarily large profit by
delaying the payment and adjusting the fixed payment accordingly. However, in practice, payments cannot be delayed forever.
If we impose a bound $\bar{K}$ on the payment date,
then the optimal solution involves setting $K=\bar{K}$ and $\phi=\frac{\varphi(\mu_0)}{\delta_{\text{DM}}^{\bar{K}}}$.
 Since the information provider cares more about future payoffs and the DM cares more about the early payoffs,
both players benefit from trading payoffs over time.\footnote{The situation in which players have different discount factors is also studied in the
repeated games literature, see, e.g., \citet{Lehrer1999}.}
Like the upfront fee, the pre-agreed lump sum payment with delay guarantees that the DM's stopping strategies are undistorted,
so that the information provider can reap the benefit of the DM making accurate decisions.

To summarize, in terms of policy implications, our results suggest that when the discount factor of the information provider is no greater than that of the
DM, charging a lump sum payment at the beginning is optimal.
However, when the information provider's discount factor is greater, then he should delay the payment
as much as possible and adjust the amount of payment at due date accordingly.

\bigskip

\headings{Timing of Payments.}
The previous analysis assumes that if the DM chooses to acquire information in period $n$,
then the fee is paid in period $n+1$ (see Eq. (\ref{Value_Func_V_n}) and Eq. (\ref{Total expected cos under c})).
A natural question is what happens if the payment is made immediately after the information acquisition decision is made and the signal is generated.
In this case, for a fee scheme $c\in C(\mu_0)$, the value function when $n\geq 1$ takes the form
\begin{equation*}
V_n^N(\mu;h_n) = \max\left\{ \pi(\mu), \delta  \E \left( V^N_{n+1}(\mu(s);h_{n+1}) | \mu; h_n \right) - \E(c(h_{n+1})|\mu; h_n)  \right\}.
\end{equation*}
Note that since the payment is made at a different time as before, the value functions are different as well.
The discounted expected total payment equals
\begin{equation*}
\rho(\mu_0;c) := c(\varnothing)+ \sum_{n=1}^{N} \sum_{h_n\in H_n} \delta^{n-1}  c(h_n) \Prob\left(h_n|\mu_0\right) \mathbbm{1}_{\left\{\mu_0(h_{n,m})\notin E^N_m(h_{n,m};c),\ \mu_0(h_n)\in A^N_n(h_n;c) \right\}}.
\end{equation*}
These expressions clearly show that a fee function $c(h_n)$ with no payment delay is equivalent to the fee scheme $\frac{1}{\delta}c(h_n)$
when the payment is made in the next period, as in our previous analysis.
Therefore, the upfront fee scheme remains optimal in this case.

More generally, we may consider other possible payment timing schemes, for instance, by allowing the payment to be delayed by a pre-determined number of periods.
But each such payment scheme is equivalent to a payment scheme with no delay, so Theorem \ref{Theorem_Compare Value of Info} remains  valid
and the value of the information structure is the same.

\bigskip

\headings{Choosing How Much Information to Acquire.} In the previous analysis, a signal is generated from the information structure every period,
and the DM decides sequentially whether to pay for the signal or not.
However, in many situations, the DM may also decide how much information to acquire (as in, for instance,
\citet{Frankel2019}, \citet{Bloedel2020} and \citet{Moscarini2001}).

To capture this feature, suppose that the information provider is endowed with a known information structure $S$,
which can generate at most $K$ conditionally independent signals per period.
We assume that due to limited resources, the information provider can generate at most $L$ conditionally independent signals from $S$ during the entire decision horizon.
The $L$ signals can be generated all at once, or sequentially, depending on the decisions of the DM. The DM chooses how many signals to purchase each time,
subject to the constraints that the number of signals purchased each period does not exceed  $K$, and  total number of signals purchased does not exceed the quota $L$.\footnote{\citet{Moscarini2001} consider a similar problem
(with time-invariant state of nature and no constraint on the total number of signals)
when the DM faces a convex cost function, which is increasing in the number of signals acquired.
Their objective is to characterize the optimal information acquisition strategy. They show that the optimal number of signals grows in the Bellman value
prior to stopping and acting.}
The information provider sets a fee scheme $c(h_n)$ that depends on the history of past signals.
Note that a history contains information on the number of signals purchased in the past as well as signal realizations.

Let $\l_n\le K$ be the number of signals the DM chooses to acquire in period $n$.
Fix a fee scheme $c$ and a history $h_n$. 
Then the period-$n$ value function can be written as
{\footnotesize{\begin{equation*}
V_n^N(\mu;h_n) =  \max \Big\{ \pi(\mu), \max_{\substack{\l_n \le K \\ \text{s.t.}\ \sum_{i=1}^{n}\l_i\leq L }} \delta \big[ \E \left( V^N_{n+1}(\mu(s_1,...,s_{\l_n});h_{n+1}) | \mu; h_n \right) - \E( c(h_{n+1})|\mu;h_n) \big]  \Big\}.
\end{equation*}
}}

\noindent Note that the history $h_{n+1}$ contains the realization of the $\l_n$ signals $s_1,...,s_{\l_n}$, and the expectations are taken with respect to
signals $s_1,...,s_{\l_n}$.
Using this recursive equation, one can show that the similar decomposition of value function (Eq. (\ref{Decompose value function})) still holds,
i.e., $V^N_0(\mu_0;\varnothing;c) = \hat{V}^N_0(\mu_0;\varnothing;\pmb{0}) - \rho(\mu_0;c)$.
But here $\hat{V}^N_0(\mu_0;\varnothing;\pmb{0})$ is the DM's discounted expected gross payoff when
she adopts the optimal information acquisition strategy (i.e., the choice of $\l_n$)
and stopping strategy corresponding to the fee function $c$,
and $\rho(\mu_0;c)$ is the discounted expected total cost as defined in Eq. (\ref{Total expected cos under c}).
Applying the same argument in the proof of Theorem \ref{Theorem_Compare Value of Info}, we conclude that the upfront fee scheme remains optimal.
The result implies that when the state of nature is time-invariant, it is optimal for the DM to acquire as much signals as possible at the beginning.
However, when the state of nature evolves over time instead of being time-invariant, the DM may benefits from postponing information acquisition.
Example 1 is an illustration.


\bigskip

\headings{Markovian State of Nature.}
Suppose the state of nature $\theta$ evolves according to a known Markov chain $\tau=(\tau_{ij})_{i,j\in\{1,...,m\}}$,
where $\tau_{ij}$ is the probability that the state changes from $\theta_i$ to $\theta_j$.
Each period, in addition to the options to stop or to acquire information,
the DM can wait without acquiring information (referred to \emph{wait} henceforth).
The DM who chose to wait previously can opt to acquire information in later periods.
The option to wait without acquiring information is relevant for two reasons:
First, with a transition matrix $\tau$, the belief may evolve gradually to
a favorable one with sufficiently high probability, even without additional information.\footnote{Consider the mapping
$\Gamma:\Delta\Theta\rightarrow\Delta\Theta$ defined by $\Gamma(\mu)=\mu\tau$.
This mapping captures the evolution of belief based solely on the transition matrix $\tau$.
If all elements of $\tau$ are strictly less than $1$, then one can show that $\Gamma$ is a contraction mapping.
Hence, it has a unique fixed point $\mu^*$, to which the belief process (without addition information) will converge.
Sometimes convergence can be achieved immediately.
For instance, if $\tau=\big(\begin{smallmatrix}  0.8 & 0.2 \\ 0.8 & 0.2 \end{smallmatrix}\big)$, then regardless of the prior, the belief next period
will be $(0.8,0.2)$. }
Second, acquiring extra information may be costly.

If the DM chooses to acquire information in period $n$,
then she would receive a noisy signal about the prevailing state in period $n+1$.\footnote{Alternatively, one may assume that if the DM chooses to
acquires information in period $n$, then she gets a signal about the period-$n$ state, instead of the period-($n+1$) state.
But in period $n+1$, the state will evolve to a new one. Although under these two approaches the beliefs held in period $n+1$ are typically different,
the qualitative results remain the same.}
Based on the transition matrix $\tau$ and the noisy signal $s$ obtained, the DM updates, and her posterior belief would be
\begin{equation*}
  \hat{\mu}(s) := \left(\frac{\hat{\mu}_{\theta}f_{\theta}(s)}{\sum_{\theta'\in\Theta}\hat{\mu}_{\theta'}f_{\theta'}(s) }\right)_{\theta\in \Theta},
\end{equation*}
where $\hat{\mu}=(\hat{\mu}_{\theta})_{\theta\in\Theta}=\mu\tau$ is the belief after accounting for the transition.
Based on the posterior belief $\hat{\mu}(s)$ in period $n+1$, the DM decides whether to stop, wait, or acquire information.

We say that a history of signals is \emph{relevant to} the DM, if it contains only the realized signals when the DM chose to acquire information and not
those when she chose to wait (these signals are unobservable by the DM).
The cost of information can be made contingent on the history of signals that is relevant to the DM.
We investigate the optimal fee scheme that depend on histories relevant to the DM.

As in Section \ref{Section: Optimal Fee Scheme}, the information provider sets a fee scheme $c(\cdot)$, the DM either takes it or leaves it.
With a Markovian state of nature, the DM's  maximal payoff if she rejects the fee scheme would be
\begin{equation*}
  \Pi(\mu_0):=\max\left\{\pi(\mu_0), \max_{n\in\{1,...,N\}}\delta^n\pi(\mu_0\tau^n)  \right\}.
\end{equation*}
If the DM accepts the fee scheme, her continuation value when she chooses to wait is $\delta V^N_1(\hat{\mu}_0;h_1;c)-c(\varnothing)$,
and her continuation value when she acquires information is
$\delta \left[ \E \left( V^N_{1}(\hat{\mu}_0(s);h_{1};c) | \mu_0; \varnothing \right) - \E(c(h_{1})|\mu_0; \varnothing) \right] - c(\varnothing)$.
In this case, the perios-$0$ value function is
\begin{equation*}
\begin{aligned}
V_0^N(\mu_0;\varnothing;c) =  \max\Big\{ & \Pi(\mu_0), ~ \delta V^N_{1}(\hat{\mu}_0;h_{1};c)-c(\varnothing), \\
                                         & \delta \left[ \E \left( V^N_{1}(\hat{\mu}_0(s);h_{1};c) | \mu_0; \varnothing \right) - \E(c(h_{1})|\mu_0; \varnothing) \right] - c(\varnothing)  \Big\}.
\end{aligned}
\end{equation*}
For period $n$, $n\geq 1$, the value function can be written as follows:
\begin{equation*}
\begin{aligned}
V_n^N(\mu;h_n;c) =  \max\Big\{ & \pi(\mu), ~ \delta V^N_{n+1}(\hat{\mu};h_{n+1};c), \\
                               & \delta \left[ \E \left( V^N_{n+1}(\hat{\mu}(s);h_{n+1};c) | \mu; h_n \right) - \E(c(h_{n+1})|\mu; h_n) \right] \Big\}.
\end{aligned}
\end{equation*}
The DM will accept a fee scheme $c(\cdot)$ if and only if $V_0^N(\mu_0;\varnothing;c)\geq \Pi(\mu_0)$.
Fix a fee scheme $c(\cdot)$, using the expressions of the value functions,
one can decompose the value function $V_0^N(\mu_0;\varnothing;c)$ in the same way as in Lemma \ref{Lemma_Decompose value function}.
It follows from the similar argument as in the proof of Theorem 1 that imposing an upfront fee $V_0^N(\mu_0;\varnothing;\pmb{0})-\Pi(\mu_0)$
is optimal (assume that $V_0^N(\mu_0;\varnothing;\pmb{0})>\Pi(\mu_0)$, otherwise the information structure has value $0$).

Compared to the time-invariant state case,
the most important difference when the state is Markovian is that the belief process is no longer a martingale.\footnote{In the
Markovian case the corresponding law is referred to as a ``$T$-martingale" (see \citet{Kohlberg1999}).
If a mapping $T: X\rightarrow X$ on a normed linear space is non-expansive,
then a sequence $\{x_n\}$ that satisfies $\E(x_{n+1}|\mathcal{F}_n)=T(x_n)$ is called a ``$T$-martingale". Here,  $T=\tau$.}
This difference implies that the optimal information acquisition strategies when the state is Markovian can be very different from
that corresponding to time-invariant state of nature.
The following example demonstrates that when the state is Markovian, the DM may benefits from postponing information acquisition.

\bigskip

\noindent\textbf{Example 1}. Consider a 3-period ($N=2$) decision problem with two states $\theta_1$, $\theta_2$, and transition matrix
$\tau=\Big(\begin{smallmatrix}  \tau_{12} & \tau_{21} \\ \tau_{21} & \tau_{22} \end{smallmatrix}\Big)$, where $\tau_{12}=0.6$, $\tau_{21}=0.4$.
The prior belief is $\mu_0=(0.1,0.9)$, where the first number is the probability that the state is $\theta_1$.
Suppose $A=\{a\}$, i.e., there is a single action, say ``invest", and the state-dependent payoff function is given by $u(a,\theta_1)=10$, $u(a,\theta_2)=-10$.
Hence invest yields positive payoff when the chance of $\theta_1$ is above $0.5$.
The DM can get information about the prevailing state from an information structure
$S=\Big(\begin{smallmatrix}  f_{\theta_1}(s_1) & f_{\theta_1}(s_2) \\ f_{\theta_2}(s_1) & f_{\theta_2}(s_2) \end{smallmatrix}\Big)$
either in the first period, or in the second period (i.e., $L$=1), where $f_{\theta_1}(s_1)=0.55$, $f_{\theta_2}(s_1)=0.45$.
\begin{itemize}
  \item If the DM chooses to get information using $S$ about the prevailing state of the first period, then depending on the signal received,
        the first period posterior belief would be either $\hat{\mu}_1(s_1)=(0.47,0.53)$, or $\hat{\mu}_1(s_2)=(0.372,0.628)$.
        Since $S$ is fully used in the first period, there is no extra capacity left in the
        second period, so the DM can only rely on the knowledge of $\tau$ to update her belief about the prevailing state in the second period.
        Therefore, the second period belief would be either $\hat{\mu}_1(s_1)\tau=(0.494,0.506)$,
        or $\hat{\mu}_1(s_2)\tau=(0.474,0.526)$.
        Hence the optimal choice involves not investing and getting 0 payoff.
  \item If the DM chooses to get information using $S$ about the prevailing state in the second period,
        then her belief in period 1 would be $\mu_0\tau=(0.42,0.58)$.
        In the second period, the belief before getting signal is $\mu_0\tau^2= (0.484,0.516)$. Given this belief,
        with probability $0.502$, signal $s_2$ will be received, the posterior would become $(0.434.0.566)$;
        with probability $0.498$, signal $s_1$ will be received, the posterior would become $(0.534,0.466)$, and it is optimal for the DM to take action $a$
        and receive a strictly positive payoff.
\end{itemize}
Hence by postponing information acquisition to the second period,
the DM receives a strictly higher expected payoff than getting information in the first period. \hfill $ \blacksquare $

\section{The Quality of Information}\label{Section: Comparing Info Structures}

In this section we consider the impact of the quality of the information that the DM receives during the decision process
on her optimal strategies.


Consider two different information structures, $S$ and $T$.
We say $S$ is more informative than $T$ (in the sense of Blackwell) and write $S\succsim T$, if $T$ can be obtained
from $S$ through a stochastic transformation.
More formally,
\begin{definition}\label{Def_More_precise_info_structure}
We say an information structure $S$ is more informative than another one $T$, and write $S\succsim T$,
if there exists a stochastic matrix $M$ such that $SM=T$.
\end{definition}

Let $\mu(s)$ and $\mu(t)$ be posterior beliefs after receiving signals $s$ and $t$ from $S$ and $T$, respectively.
\citet{Blackwell1953} shows that the following condition is equivalent to $S\succsim T$:
for \emph{every} convex and continuous function $h: \Delta\Theta\rightarrow \mathbb{R}$,
$\E_S(h(\mu(s)))\geq\E_T(h(\mu(t)))$.
Related to our paper, one can show that the value functions are convex and continuous,
hence Blackwell's equivalence theorem can be applied to compare value functions under information structures of different quality.

\ignore{
Denote by $\alpha_s(\mu)$ the probability of receiving signal $s$ under $S$ when the belief is $\mu$, and by $\beta_t(\mu)$
the probability of receiving signal $t$ under $T$.\footnote{When the belief is $\mu$, $\mu=(\mu_\theta)_{\theta\in\Theta}$,
$\alpha_s(\mu)=\sum_{\theta\in\Theta} \mu_\theta f_{\theta}(s)$, $\beta_t(\mu)=\sum_{\theta\in\Theta} \mu_\theta g_{\theta}(t)$.
Here $f_\theta(s)$ and $g_\theta(t)$ are the probabilities of receiving signals $s$ and $t$
when the true state is $\theta$ under the information structures $S$ and $T$, respectively.  }
\citet{Blackwell1953} shows that the following condition is equivalent to $S\succsim T$:
for \emph{every} convex and continuous function $h: \Delta\Theta\rightarrow \mathbb{R}$,
$\sum_{s}\alpha_s(\mu) h(\mu(s))\geq \sum_{t}\beta_t(\mu) h(\mu(t))$.
The condition states that $S$ is more informative than $T$ if and only if for every convex continuous function $h$,
the expected value of $h$ under $S$ is greater than that under $T$.
}

\begin{proposition} \label{Prop_V_n_convex}
Given any history $h_n$, the value function $V_n^N(\mu;h_n)$ and the conditional expectation of the value function
$\E(V_{n+1}^N(\mu(s);h_{n+1})|\mu;h_n)$ are convex and continuous in the belief $\mu$.
\end{proposition}

In what follows, we consider information structures $S$ and $T$ with $S\succsim T$.
Recall from Theorem \ref{Theorem_Compare Value of Info} that the value can be achieved via an upfront fee, and that the optimality
of the upfront fee scheme does not depend on the information structure. Therefore, we restrict attention to upfront fee schemes.
From a decision making perspective, a posterior belief is a sufficient statistic for the past history of signals.
We therefore regard the value functions as functions of posterior beliefs.
Let $\{A^N_n(S)\}_{0\leq n\leq N}$ and $\{A^N_n(T)\}_{0\leq n\leq N}$
be the corresponding optimal information acquisition sets and let
$\{E^N_n(S)\}_{0\leq n\leq N}$ and $\{E^N_n(T)\}_{0\leq n\leq N}$ be the optimal stopping sets.
The following result highlights the impact of information quality on the value function and optimal strategies.  
\begin{proposition}\label{Theorem_compare entry sets}
Consider a stopping problem. Let  $S$ and $T$ be information structures with $S\succsim T$.
Denote  by $V^N_n(\mu)$ and $U^N_n(\mu)$ the period-$n$ value functions when the \emph{DM} receives information from $S$ and $T$, respectively. Then,
\begin{enumerate}
  \item For any period $n$, $V^N_n(\mu)\geq U^N_n(\mu)$, $\forall \mu\in\Delta\Theta$; 
  \item  $A^N_n(T)\subseteq A^N_n(S)$, $E^N_n(S)\subseteq E^N_n(T)$.
\end{enumerate}
\end{proposition}

Proposition \ref{Theorem_compare entry sets} can be easily extended to the infinite-horizon case ($N=\infty$). 
Assume that $\delta<1$ and  that the DM's payoff when she never stops is 0.
Let the value functions in the infinite-horizon problem under information structures $S$ and $T$ be $V^\infty(\mu)$ and $U^\infty(\mu)$, respectively.
(Note that $V^\infty(\mu)$ and $U^\infty(\mu)$ do not account for the upfront fee that is paid immediately
in period 0 if the DM chooses to wait for at least one period.)
The function $V^\infty(\mu)$ satisfies the functional equation
\begin{equation}\label{V_Bellman_Equ_Infiinte}
V^\infty(\mu) = \max\left\{\pi(\mu), \delta \E(V^\infty(\mu(s))|\mu)\right\}.
\end{equation}
One can show that 
the functional Eq.\ (\ref{V_Bellman_Equ_Infiinte}) is a contraction mapping.
Hence, as stated in the following lemma, $V^\infty(\mu)$ is the limit of the sequence of value functions as the decision horizon goes to infinity.
\begin{lemma}\label{Lemma_Infitnite horizon}
Suppose $\delta<1$. There exists a unique function $V^\infty(\mu)$ that satisfies Eq.\ \emph{(\ref{V_Bellman_Equ_Infiinte})}.
Moreover, $V^\infty(\mu)=\lim_{N\rightarrow \infty} V_0^N(\mu)$. 
\end{lemma}

By Proposition \ref{Theorem_compare entry sets}, $V^N_0(\mu)\geq U^N_0(\mu)$ for all $N$.
In the limit, we have $V^\infty(\mu)\geq U^\infty(\mu)$,
and this implies that the optimal waiting set corresponding to $S$ is larger in the infinite-horizon case.

Proposition \ref{Theorem_compare entry sets} says that the value functions under a Blackwell-dominating information structure $S$ are uniformly greater.
This implies that in every period, the set of beliefs that lead to waiting is larger under a Blackwell-dominating information structure.
The following result shows that having a greater discount factor has the same effect in terms of optimal strategies. 
\begin{proposition}\label{Prop_Waiting set and delta}
Consider two discount factors $\delta$ and $\hat\delta$ with $1\geq \delta>\hat{\delta}$.
Let $A^N_n(\delta),\ A^N_n(\hat{\delta})$ be the corresponding waiting sets (similarly for the stopping sets).
Then $E^N_n(\delta)\subseteq E^N_n(\hat{\delta})$,
$A^N_n(\hat{\delta}) \subseteq A^N_n(\delta)$.
The stopping time distribution induced by $\delta$ first-order stochastically dominates that induced by $\hat{\delta}$.
\end{proposition}

The last part of Proposition \ref{Prop_Waiting set and delta} implies that the expected stopping time corresponding to a greater discount factor is longer.
A natural question is whether higher information quality has a similar implication in terms of stopping time.
Note that the expected stopping time is affected by two opposite forces. 
On the one hand, larger waiting sets under the more informative information structure may contribute to longer expected stopping time.
On the other hand, signals generated from the higher quality information structure tend to be more informative and contribute to faster belief updating.
Clearly, there is a trade-off.
Which force prevails depends on the problem under consideration.

\bigskip

\begin{figure}[htp]
\centering
\includegraphics[scale=0.42]{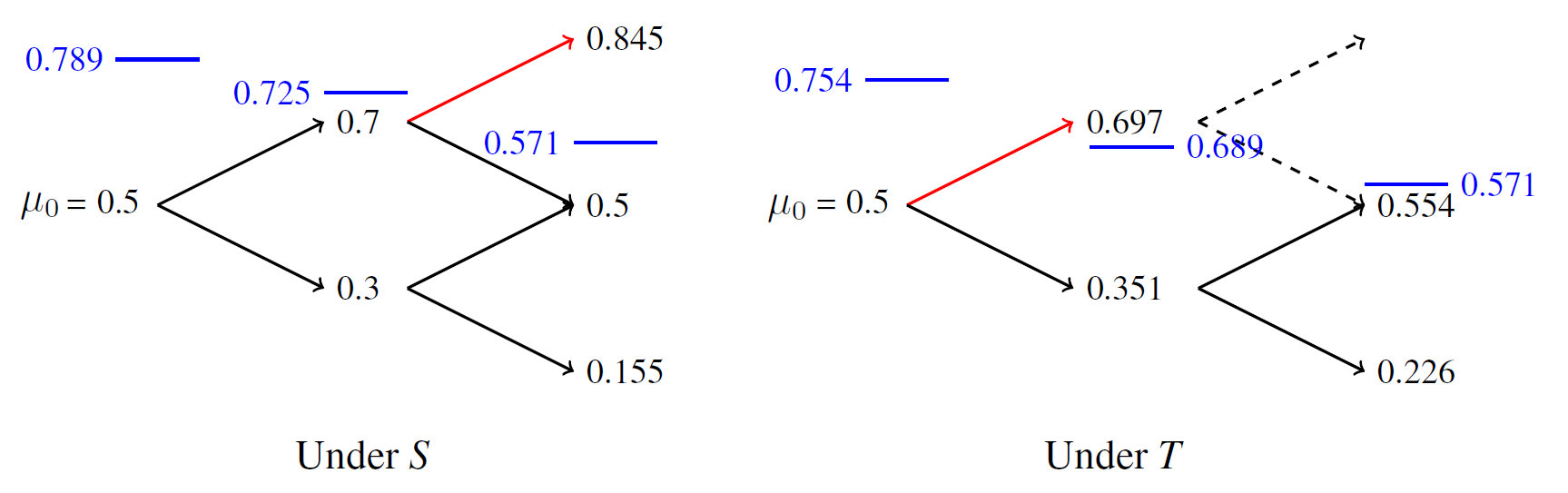}
\caption{Comparison of expected stopping time -- first example}
\label{fig_waiting time example 1}
\end{figure}

\ignore{
{\color{blue} Tao: I think Example 1 below should be removed, since Example 2 below already illustrates the trade off on expected stopping time and Example 2
is more informative than a numerical example like Example 1. } \Red{I let you decide. I think that this example cannot cause any harm. Furthermore, it is a reminiscent from the previous version - to show that we revised the paper and didn't actually submit a new one. } }
\noindent\textbf{Example 2}.  Consider a 3-period ($N=2$) stopping problem with time-invariant state of nature $\theta$,
$\theta\in\{\theta_1, \theta_2\}$. The action space contains a single action $A=\{a\}$,
the payoff function is $u(a,\theta_1)=6,\ u(a,\theta_2)=-8$ and $\delta=0.9$.
Let $\mu_0=0.5$ be the prior belief that the state is $\theta_1$.\footnote{At the risk of abusing
notations, when there are two states, we use a single number $\mu$ to represent the belief. }
Suppose $S=\Big(\begin{smallmatrix}  f_{\theta_1}(s_1) & f_{\theta_1}(s_2) \\ f_{\theta_2}(s_1) & f_{\theta_2}(s_2) \end{smallmatrix}\Big)$,
where $f_{\theta_1}(s_1)=0.7$,  $f_{\theta_2}(s_1)=0.3$,
and $T=\Big(\begin{smallmatrix}  g_{\theta_1}(t_1) & g_{\theta_1}(t_2) \\ g_{\theta_2}(t_1) & g_{\theta_2}(t_2) \end{smallmatrix}\Big)$,
where $g_{\theta_1}(t_1)=0.6$, $g_{\theta_2}(t_1)=0.261$.
It can be easily verified that $S\succsim T$. 
The optimal stopping thresholds are represented by the ``bars" in Figure \ref{fig_waiting time example 1}.
The arrows indicate the evolution of the DM's beliefs. Once beliefs cross the ``bars", it is optimal for the DM to stop.
In this example, although the ``good" signal under the more informative information structure $S$ has greater strength, in the sense that
it contributes to faster belief updating (one can simply compare the likelihood ratios of $s_1$, $t_1$: $0.7/0.3>0.6/0.261$),
the effect of higher thresholds prevails. As a result, the expected stopping time when DM anticipates to receive information from $S$ is longer.

\begin{figure}[htp]
\centering
\includegraphics[scale=0.42]{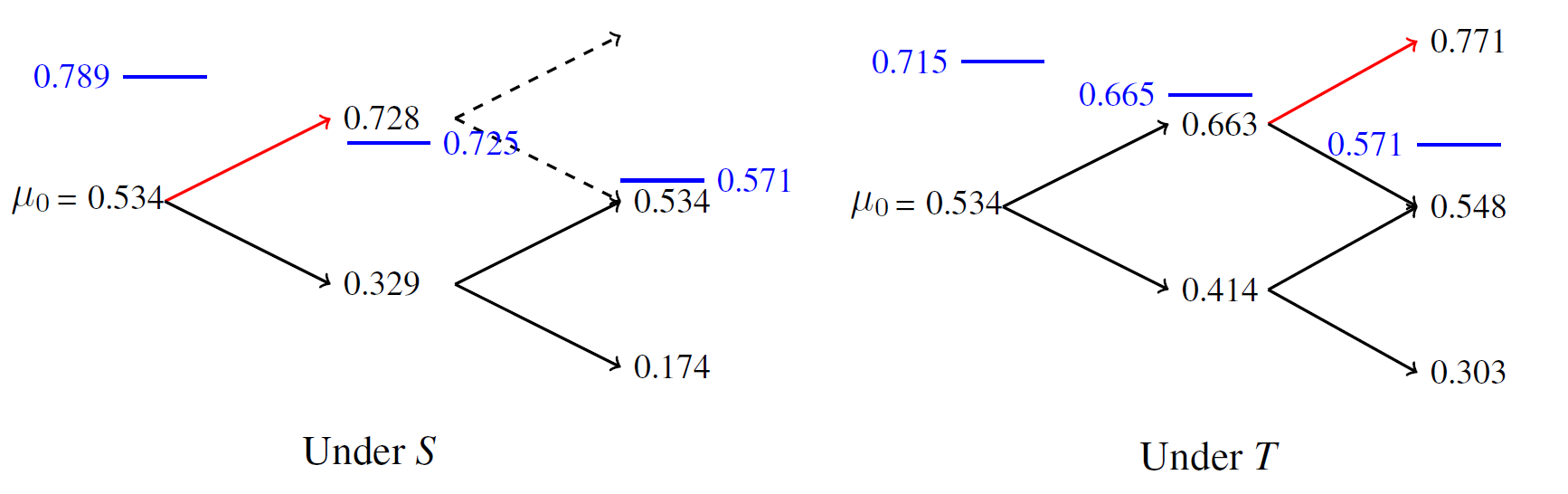}
\caption{Comparison of expected stopping time -- second example}
\label{fig_waiting time example 2}
\end{figure}

Now consider the same decision problem, but with a different information structure $T$,
$T=\Big(\begin{smallmatrix}  g_{\theta_1}(t_1) & g_{\theta_1}(t_2) \\ g_{\theta_2}(t_1) & g_{\theta_2}(t_2) \end{smallmatrix}\Big)$,
where $g_{\theta_1}(t_1)=0.6$, $g_{\theta_2}(t_1)=0.35$. $S$ remains the same.
Clearly, $S\succsim T$.
Again the optimal stopping thresholds are represented by the ``bars" in Figure \ref{fig_waiting time example 2} and the arrows indicate the evolution of beliefs.
In contrast to the previous example, the expected stopping time is shorter under the more informative information structure $S$,
because the effect of greater strength of the ``good" signal under $S$ dominates the effect of higher stopping thresholds.  \hfill $ \blacksquare $

\ignore{
$S=\left(
    \begin{array}{cc}
      0.7 & 0.3 \\
      0.3 & 0.7 \\
    \end{array}
  \right)\ \text{and} \
T=\left(
    \begin{array}{cc}
      0.6          & 0.4  \\
      0.261        & 0.739 \\
    \end{array}
  \right).$
$S=\left(
    \begin{array}{cc}
      0.7 & 0.3 \\
      0.3 & 0.7 \\
    \end{array}
  \right) \text{and} \
T=\left(
    \begin{array}{cc}
      0.6   & 0.4  \\
      0.35  & 0.65 \\
    \end{array}
  \right).$  }

\bigskip

Generally speaking, calculating the expected stopping time can be very difficult.
In some special situations, however, we can obtain a closed-form expression for the expected stopping time.
We use such situations to illustrate the trade-off on expected stopping time discussed above.

\bigskip

\noindent\textbf{Example 3} [Normally distributed signals and expected stopping time].
We now consider the case where the stopping time under both $S$ and $T$ are unbounded.
Suppose $\theta\in\{\bar{\theta},\underline{\theta}\}$,
$\bar{\theta}$, $\underline{\theta}\in\mathbb{R}$, with  $\bar{\theta}>\underline{\theta}$, and the time horizon is infinite.
Assume the action set $A$ contains a single action, say invest.
Given an information structure, the optimal waiting/stopping strategies in this case can be characterized by a unique threshold of belief:
once the belief that the state is $\bar\theta$ reaches the threshold, it is optimal to stop, otherwise, it is optimal to wait.
Consider two information structures $S=\left(f_\theta(s)\right)_{\theta\in\Theta}$
and $T=\left(g_\theta(t)\right)_{\theta\in\Theta}$, where $f_\theta(s)\sim N(\theta,\sigma_S^2)$,
and $g_\theta(t)\sim N(\theta,\sigma_T^2)$, with $\sigma_T>\sigma_S$.
Conditional on each state $\theta$, the probability distribution $g_\theta$ has the same mean as $f_\theta$, but the former has larger variance.
It follows that $S\succsim T$.\footnote{Note that Definition \ref{Def_More_precise_info_structure}, which applies to discrete signal spaces,
can be generalized as follows. Let $s$ and $t$ be random signals under $S$ and $T$, respectively. Then $S\succsim T$ if there exit a random signal $z$
with known distribution and a function $h$ such that for every $\theta$, $h(s,z)$ is distributed as $g_\theta$.
For normal distributions as in our case, $h(s,z)=s+z$ and $z$ is a normal random variable with mean $0$. See, e.g., \citet{Lehmann1988}.}
Let $\bar{\mu}_S$ and $\bar{\mu}_T$ be the stopping thresholds under $S$ and $T$, respectively.
Proposition \ref{Theorem_compare entry sets} and Lemma \ref{Lemma_Infitnite horizon} imply that $\bar{\mu}_S\geq \bar{\mu}_T$.
In what follows, we focus on the generic case in which $\bar{\mu}_S>\bar{\mu}_T$ and compare the expected stopping time.

To make the problem non-trivial, suppose $\mu_0<\bar{\mu}_S$. Define $r_0:=\ln \frac{\mu_0}{1-\mu_0}$ as the log-likelihood ratio of the prior.
Let $\mu(s_1,...,s_n)$ and $r(s_1,...,s_n)$ be the posterior belief and log-likelihood ratio after receiving a sequence of signals from $S$
(similarly, from $T$), respectively. By Bayes rule,
\begin{eqnarray*}
\frac{\mu(s_1,...,s_n)}{1-\mu(s_1,...,s_n)} &=& \frac{\mu_0}{1-\mu_0} \frac{f_{\bar{\theta}}(s_1,...,s_n)}{f_{\underline{\theta}}(s_1,...,s_n)} \\
                                            &=& \frac{\mu_0}{1-\mu_0} \exp\left(\sum_{i=1}^{n}\frac{(s_i-\bar{\theta})^2-(s_i-\underline{\theta})^2}{2\sigma_S^2}\right).
\end{eqnarray*}
It follows that
\begin{equation*}
  r(s_1,...,s_n) = r_0 + \frac{1}{\sigma_S^2}(\bar{\theta}-\underline{\theta})\sum_{i=1}^n\left(s_i-\frac{\bar{\theta}+\underline{\theta}}{2}\right).
\end{equation*}
The advantage of dealing with the log-likelihood ratio of the prior is made clear by this formula: it is linear in $s_i$.
If we write $x_i:=\frac{1}{\sigma_S^2}(\bar{\theta}-\underline{\theta})\left(s_i-\frac{\bar{\theta}+\underline{\theta}}{2}\right)$,
then conditional on a given state $\theta$, the likelihood ratio is the sum of independent increments: $r(s_1,...,s_n)=r_0+\sum_{i=1}^{n}x_i$.
Since given $\bar\theta$, $s_i\sim N(\bar\theta,\sigma_S^2)$,
we obtain $x_i\sim N\left(\frac{1}{2\sigma_S^2}(\bar{\theta}-\underline{\theta})^2, \frac{1}{\sigma_S^2}(\bar{\theta}-\underline{\theta})^2 \right)$.
Similarly, given state $\underline{\theta}$,
$x_i\sim N\left(-\frac{1}{2\sigma_S^2}(\bar{\theta}-\underline{\theta})^2, \frac{1}{\sigma_S^2}(\bar{\theta}-\underline{\theta})^2 \right)$.

The DM will wait until the log-likelihood ratio of posterior belief hits the cutoff $\bar{r}_S:=\ln\frac{\bar{\mu}_S}{1-\bar{\mu}_S}$
(the cutoff $\bar{r}_T$ is similarly defined) for the first time.
Let $\eta_S$ and $\eta_T$ be the (random) stopping time under $S$ and $T$.
We compare the expected stopping time $\E(\eta_S)$ and $\E(\eta_T)$ conditional on $\theta=\bar{\theta}$.
By Wald's identity, $\E(r(s_1,...,s_{\eta_S}))-r_0 = \E(x_i)\E(\eta_S)$, hence
\begin{equation*}
\E(\eta_S) = \frac{2\sigma_S^2\left(\E(r(s_1,...,s_{\eta_S}))-r_0\right)}{(\bar{\theta}-\underline{\theta})^2},\ \ \text{and}\ \ \ \E(\eta_T) = \frac{2\sigma_T^2\left(\E(r(t_1,...,t_{\eta_T}))-r_0\right)}{(\bar{\theta}-\underline{\theta})^2}.
\end{equation*}
Note that since $\bar{r}_S>\bar{r}_T$, the expected stopped log-likelihood ratios satisfy $\E(r(s_1,...,s_{\eta_S}))-r_0>\E(r(t_1,...,t_{\eta_T}))-r_0$.
This implies that other things being equal, greater stopping thresholds tend to prolong the expected stopping time.
But, on the other hand, the variance under $S$ is smaller (the information quality of $S$ is higher), which tends to shorten the expected stopping time.
The expressions of $\E(\eta_S)$ and $\E(\eta_T)$ show that there exists a unique threshold of prior belief $\bar{\mu}_0$, such that
$\E(\eta_T)>\E(\eta_S)$ if $\mu_0<\bar{\mu}_0$, and $\E(\eta_T)<\E(\eta_S)$ if $\bar{\mu}_0 < \mu_0 < \bar{\mu}_S$.
Hence, when the prior is not far below the stopping threshold, the expected stopping time under a more informative information structure is longer conditional on
the state $\bar{\theta}$; but when then prior belief is sufficiently pessimistic, having access to better information tends to shorten the expected stopping time.
However, when the state is $\underline{\theta}$, the expected stopping time under both $S$ and $T$ are unbounded. \hfill $ \blacksquare $

\bigskip

\noindent\textbf{Example 4} [Random walk and stopping time distribution].
In this example, we examine the stopping time distribution and investigate whether it is true that the stopping time distribution
under a Blackwell-dominating information structure second-order stochastically dominates that under a Blackwell-inferior information structure.
Consider $A=\{a\}$, $\Theta=\{\bar{\theta}, \underline{\theta}\}$. Also, assume binary signals:
$S=\Big(\begin{smallmatrix} f_{\bar{\theta}}(s_1)   & f_{\bar{\theta}}(s_2) \\ f_{\underline{\theta}}(s_1) & f_{\underline{\theta}}(s_2) \end{smallmatrix}\Big)$,
$T=\Big(\begin{smallmatrix} g_{\bar{\theta}}(t_1)   & g_{\bar{\theta}}(t_2) \\ g_{\underline{\theta}}(t_1) & g_{\underline{\theta}}(t_2) \end{smallmatrix}\Big)$,
where $f_{\bar{\theta}}(s_1) = f_{\underline{\theta}}(s_2)=p_S,\  g_{\bar{\theta}}(t_1) = g_{\underline{\theta}}(t_2)=p_T,$ and  $p_S>p_T>\frac{1}{2}$.
Signals $s_1$ and $t_1$ lead to upward belief update, while $s_2$ and $t_2$ update belief downward.
Let $\mu_0$ be the prior belief that the state is $\bar{\theta}$, $\mu_0\in(0,1)$.
For a fixed state, the belief process (in terms of log-likelihood ratio) induced by an information structure is a random walk.

Under $\bar{\theta}$ the drift of the random walk is positive, while under $ \underline{\theta}$ it is negative.
When the decision horizon is infinite, the optimal strategy can be characterized by a time-invariant belief thresholds ($\bar{\mu}_S$, $\bar{\mu}_T$, respectively).
Consider the situation in which starting from the same prior belief, both thresholds $\bar{\mu}_S$, $\bar{\mu}_T$ can be reached in one step.
That is, once the number of $s_1$ (resp. $t_1$) signals exceeds the number of $s_2$ (resp. $t_2$) signals for the first time,
the DM decides to stop.\footnote{One can always get such a situation
by adjusting the parameters, for instance, by making $p_S$ and $p_T$ closer.}
The expected stopping time under both $S$ and $T$ is infinity, since by Wald's identity, conditional on $\underline{\theta}$,
the expected stopping time is infinity.
Denote by $\P_S(n|\theta)$ and $\P_T(n|\theta)$ the probabilities that the stopping occurs in period $n$ under $S$ and $T$, respectively,
conditional on $\theta$. 
It follows from well-known results of random walk that
the stopping time distribution takes the form\footnote{Recall the double factorial notation: $(2n-1)!! = (2n-1)(2n-3)\cdots 3\cdot 1 $.
By convention, when $n=1$, set $(-1)!!=1$. }
\begin{equation}\label{Eq_distribution of hitting time}
\P_S(2n-1|\bar{\theta}) = \frac{(2n-3)!!}{n!} 2^{n-1} p_S^n (1-p_S)^{n-1},
\end{equation}
and $\P_S(2n|\bar{\theta}) = 0$ (the threshold can only be reached in odd periods).
When conditioning on $\underline{\theta}$, just switch $p_S$ and $1-p_S$.
Unconditional on the state, the probability distribution of stopping time is given by
$\P_S(n)=\mu_0 \P_S(n|\bar{\theta})+(1-\mu_0)\P_S(n|\underline{\theta})$. The expression for $\P_T(n)$ is similar.

Let $\mathcal{P}_S$ and $\mathcal{P}_T$ be the corresponding CDFs for the stopping time distributions $\P_S(n)$ and $\P_T(n)$, respectively.
We know that $\mathcal{P}_S$ second-order stochastically dominates $\mathcal{P}_T$ if and only if for any $x\geq 0$,
$\int_{0}^{x}\mathcal{P}_T \geq \int_{0}^{x}\mathcal{P}_S$.
We use this property and Eq.\ (\ref{Eq_distribution of hitting time}) to check whether $\mathcal{P}_S$ second-order stochastically dominates $\mathcal{P}_T$.

\begin{figure}[htp]
\centering
\includegraphics[scale=0.6]{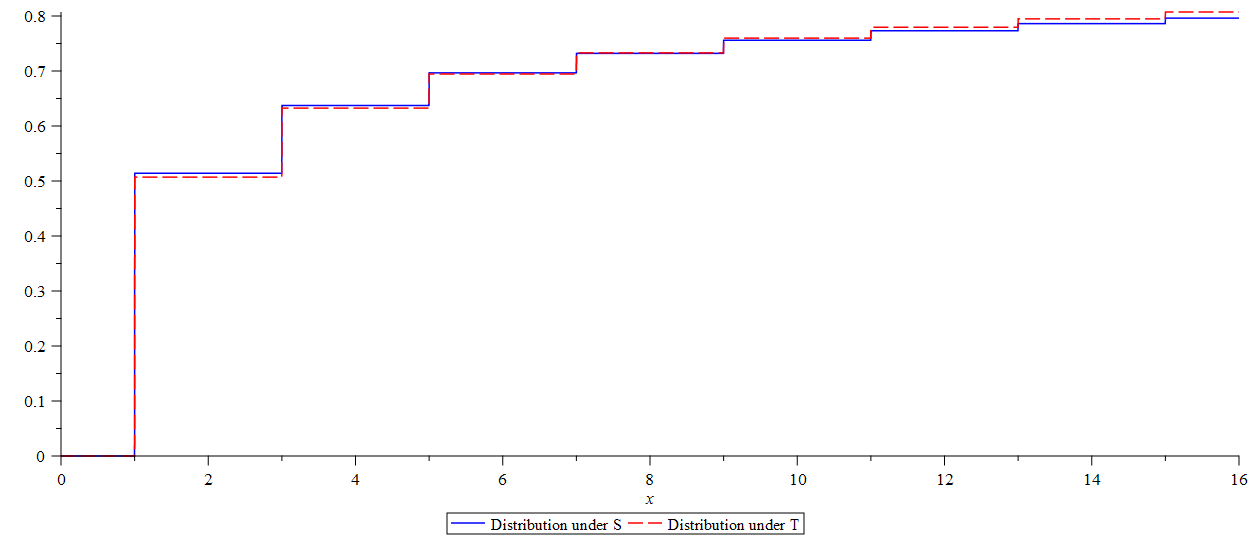}
\caption{CDFs of stopping time distribution}
\label{fig_Stopping_Time_Distribution}
\end{figure}

Let us compare $\P_S(1)$ with $\P_T(1)$.
Clearly, $\P_S(1) = \mu_0 \P_S(1|\bar{\theta})+(1-\mu_0)\P_S(1|\underline{\theta}) = \mu_0 p_S +(1-\mu_0)(1-p_S) = 1-p_S-\mu_0+2\mu_0p_S$.
Similarly, $\P_T(1)= 1-p_T-\mu_0+2\mu_0p_T$. So $\P_S(1) - \P_T(1)= (p_S-p_T)(2\mu_0-1)$.
One can ensure that $\P_S(1)>\P_T(1)$ by choosing parameters such that $\bar{\mu}_S$, $\bar{\mu}_T$ can be reached in one step and that $\mu_0>\frac{1}{2}$.
For instance, consider $\delta=0.9,\ u(a,\bar{\theta})=100,\ u(a,\underline{\theta})=-100$, $\mu_0=0.57$, and the information structures
$S=\big(\begin{smallmatrix}  0.6 & 0.4 \\ 0.4 & 0.6 \end{smallmatrix}\big)$,  $T=\big(\begin{smallmatrix} 0.55   & 0.45 \\ 0.45  & 0.55 \end{smallmatrix}\big)$.
Approximately, $\bar{\mu}_S\approx 0.66$, $\bar{\mu}_T\approx0.59$. Under both $S$ and $T$, the thresholds can be reached in one step starting from $\mu_0$.
This means for $x\in(1,2)$,  $\int_{0}^{x}\mathcal{P}_T \geq \int_{0}^{x}\mathcal{P}_S $ does not hold,
hence  $\mathcal{P}_S$ does not second-order stochastically dominates $\mathcal{P}_T$.
Figure \ref{fig_Stopping_Time_Distribution} illustrates the CDFs of the stopping time distributions in this case.

\ignore{
A related paper \citet{Zhong2017} considers the stopping time distribution induced by different learning strategies that implement the same distribution
over posterior beliefs, subject to a flow information cost constraint.
The author makes the insightful observation that under a convex (in time) discount function,
the stopping time distribution induced by the Poisson learning strategy (i.e., the DM learns the state perfectly with a Poisson process) is second-order
stochastically dominated by any other feasible learning strategies.
It follows from the convexity of the discount function that the Poisson learning process leads to the highest expected decision utility.
In comparison, we consider stopping time distributions induced by information structures of different quality.
The example shows that a similar second-order stochastic dominance relation does not hold in our setup.   \hfill $ \blacksquare $
}

We conclude this chapter by two additional remarks about Proposition \ref{Theorem_compare entry sets}.

\noindent\textbf{Remark 1}. Note that in the infinite-horizon case,
a greater value function does not imply that the underlying information structure is more informative.
To see this, suppose $\Theta=\{\underline{\theta},\bar{\theta}\}$ and the action set $A$ upon stopping consists of a single element with
$u(a,\bar{\theta})=\bar{v}>0$, $u(a,\underline{\theta})=\underline{v}<0$. Let $\mu$ be the belief that $\theta=\bar{\theta}$.
Moreover, suppose both $S$ and $T$ have binary signals:
$S=\Big(\begin{smallmatrix} f_{\bar{\theta}}(s_1)   & f_{\bar{\theta}}(s_2) \\ f_{\underline{\theta}}(s_1) & f_{\underline{\theta}}(s_2) \end{smallmatrix}\Big)$ and
$T=\Big(\begin{smallmatrix} g_{\bar{\theta}}(t_1)   & g_{\bar{\theta}}(t_2) \\ g_{\underline{\theta}}(t_1) & g_{\underline{\theta}}(t_2) \end{smallmatrix}\Big)$.
Assume the two signals $s_1$ and $t_1$ lead to upward belief updating and the strength of $s_1$
is marginally smaller than the strength of $t_1$, whereas the strength of $s_2$ is much greater than that of $t_2$.
More precisely, assume that $f_{\bar{\theta}}(s_1)/f_{\underline{\theta}} (s_1)<g_{\bar{\theta}}(t_1)/g_{\underline{\theta}}(t_1)$
and $f_{\underline{\theta}}(s_2)/f_{\bar{\theta}}(s_2)\gg g_{\underline{\theta}}(t_2)/g_{\bar{\theta}}(t_2)$.

\ignore{\footnote{Examples are abundant, for instance,
consider $S=\left(
    \begin{array}{cc}
      0.98 & 0.02 \\
      0.49 & 0.51
    \end{array}
  \right)\ \text{and} \
T=\left(
    \begin{array}{cc}
      0.42 & 0.58 \\
      0.2  & 0.8
    \end{array}
  \right).$ }
  }
Since $f_{\bar{\theta}}(s_1)/f_{\underline{\theta}}(s_1)<g_{\bar{\theta}}(t_1)/g_{\underline{\theta}}(t_1)$,
in each period, the value function under $T$ becomes positive at a lower level of belief $q$ than that under $S$ (see Panel (a)
of Figure \ref{Fig_Value and Info Structure}).
However, as the time horizon goes to infinity,
the limiting value function under $S$ is uniformly greater than the limiting value function
under $T$ (Panel (b) of Figure \ref{Fig_Value and Info Structure}).\footnote{This is because the sequence of belief levels
at which the value functions become positive converges to 0 under both $S$ and $T$. }
But it is clear that $S$ and $T$ are not Blackwell comparable.

\begin{figure}[h!]
  \centering
  \begin{tikzpicture}[scale=1]

  \draw[line width=0.3mm,->] (0,0) node[left,black]{\small 0} --(0,5);
  \draw[line width=0.3mm,->] (0,0)--(5,0) node[below,black]{$\mu$};

  \draw[line width=0.5mm] (2,0)--(4,4);
  \draw[gray,dashed,line width=0.3mm] (4,4)--(4,0) node[below,black] {1};

  \draw[line width=0.3mm, red] (1,0)--(2.5,1);     \draw[line width=0.5mm, red] (2.5,1)--(4,4);
  \draw[line width=0.3mm, blue] (0.8,0)--(2.2,0.4);    \draw[line width=0.4mm, blue] (0,0)--(0.8,0);   \draw[line width=0.5mm, blue] (2.2,0.4)--(2.5,1);

  \draw[line width=0.2mm,->] (2,0.7)--(1.5,1.3) node [above,black]{$V^1_0(\mu)$};
  \draw[line width=0.2mm,->] (1.8,0.3) to [out=-30,in=-135] (3,0.5)node [right,black]{$U^1_0(\mu)$};
  \draw[line width=0.2mm,->] (3.47,3)--(2.5,4) node [above,black]{$\mu\bar{v}+(1-\mu)\underline{v}$};

  \node at (2.5,-1.2) {\text{Panel (a): Value functions $V^1_0(\mu)$ and $U^1_0(\mu)$}};

  \draw[line width=0.3mm,->] (8,0) node[left,black]{\small 0} --(8,5);
  \draw[line width=0.3mm,->] (8,0)--(13,0) node[below,black]{$\mu$};

  \draw[line width=0.5mm] (10,0)--(12,4);
  \draw[gray,dashed,line width=0.3mm] (12,4)--(12,0) node[below,black] {1};

  \draw[line width=0.3mm, red] (8,0) to [out=8,in=-120] (11.7,3.4);   \draw[line width=0.5mm, red] (11.7,3.4)--(12,4);
  \draw[line width=0.3mm, blue] (8,0) to [out=2,in=-120] (10.7,1.4);  \draw[line width=0.5mm, blue] (10.7,1.4)--(11.7,3.4);

  \draw[line width=0.2mm,->] (11.75,3.5)--(10.8,4.2) node [above,black]{$\mu\bar{v}+(1-\mu)\underline{v}$};
  \draw[line width=0.2mm,->] (10.5,1.5)--(10,2) node [above,black]{$V^\infty(\mu)$};
  \draw[line width=0.2mm,->] (10,0.5) to [out=-30,in=-135] (11,0.8)node [right,black]{$U^\infty(\mu)$};

  \node at (10.5,-1.2) {\text{Panel (b)}: The limit case};

 \end{tikzpicture}
\caption{$V^\infty(\mu)\geq U^\infty(\mu)$ does not imply $S\succsim T$}
\label{Fig_Value and Info Structure}
\end{figure}

\bigskip

\noindent\textbf{Remark 2}. The converse of the second part of Proposition \ref{Theorem_compare entry sets} may not hold.
As the following example illustrates, it may happen that the optimal waiting set under $S$ in each period is larger
than that under $T$, and yet $S\not \succsim T$.

\bigskip

\noindent\textbf{Example 5}. In this example we show that it could be that larger waiting sets under
$S$ than those under $T$ imply neither $S\succsim T$ nor $T\succsim S$. Consider a problem where $\Theta=\{\bar\theta, \underline{\theta}\}$, $A=\{a\}$,
$u(a,\bar{\theta})=6,\ u(a,\underline{\theta})=-8,\ \delta=0.85,\ N=5$. Let $\mu$ be the belief that $\theta=\bar{\theta}$.
The optimal waiting set in each period can be characterized by a unique threshold $\bar{\mu}^N_n$ in $[0,1]$.
Consider the  information structures
$S=\Big(\begin{smallmatrix} f_{\bar{\theta}}(s_1)   & f_{\bar{\theta}}(s_2) \\ f_{\underline{\theta}}(s_1) & f_{\underline{\theta}}(s_2) \end{smallmatrix}\Big)$ and
$T=\Big(\begin{smallmatrix} g_{\bar{\theta}}(t_1)   & g_{\bar{\theta}}(t_2) \\ g_{\underline{\theta}}(t_1) & g_{\underline{\theta}}(t_2) \end{smallmatrix}\Big)$,
where $f_{\bar{\theta}}(s_1)= 0.8$, $f_{\underline{\theta}}(s_1) = 0.5$, $g_{\bar{\theta}}(t_1) = 0.6$ and $g_{\underline{\theta}}(t_1)=0.3$.
The result of  comparing the optimal stopping thresholds under $S$ and $T$ is summarized in Table \ref{Table 1_Example}.

\bigskip

\begin{table}[ht]
\centering 
\begin{tabular}{c c c c c c c} 
\hline\hline 
 & $n=N=5$ & $n=4$ & $n=3$ & $n=2$ & $n=1$ & $n=0$ \\ [0.5ex] 
\hline 
$\bar{\mu}_n^N(S)$                & 0.5714 & 0.7055  & 0.7201  & 0.7335  & 0.7335  & 0.7376  \\ 
$\bar{\mu}_n^N(T)$                & 0.5714 & 0.6697  & 0.7142  & 0.7225  & 0.7263  & 0.7312  \\
$\bar{\mu}_n^N(S)-\bar{\mu}_n^N(T)$ &      0 & 0.0358  & 0.0059  & 0.0110  & 0.0072  & 0.0064  \\ [1ex] 
\hline 
\end{tabular}\caption{Optimal stopping thresholds under $S$ and $T$}
\label{Table 1_Example} 
\end{table}
Here, the optimal stopping thresholds are uniformly higher under $S$ than under $T$ (i.e., the waiting sets under
$S$ are larger than those under $T$). However, neither $S\succsim T$, nor $T\succsim S$.

\ignore{
Let $\rho^*(\mu_0;S)$, $\rho^*(\mu_0;T)$ be the values of information structures $S$ and $T$, respectively.
An immediate implication of Theorem \ref{Theorem_Compare Value of Info} and Theorem \ref{Theorem_compare entry sets} is as follows.
\begin{corollary}\label{Corollary: comparison of values}
If $S\succsim T$, then for any prior belief $\mu_0\in\Delta\Theta$, $\varphi(\mu_0;S)\geq\varphi(\mu_0;T)$.
\end{corollary}
So the value of $S$ in every dynamic entry problem is greater than that of $T$ whenever $S$ is more informative than $T$ in the sense of Blackwell.
}

\section{Conclusion}\label{Section_conclusion}
In this paper we consider a wide class of sequential decision problems in which a DM faces a payoff-relevant unknown state of nature,
and decides sequentially whether to stop or wait and get information from a noisy information structure.
Unlike the previous literature that studies the value and optimal strategies under an exogenously given information cost function,
we study the optimal choice of such cost functions and the value of a given information structure in such stopping problems.

We show that among all history-dependent information cost functions, the upfront fee scheme is optimal,
in the sense that it generates the highest discounted expected total payoff for the owner of the information structure (information provider).
The optimal upfront fee is exactly the value of the information structure.
Additionally, the optimality of the upfront fee does not hinge on the information structure under consideration.
Our result suggests that the widely adopted subscription fee payment method for many private databases and consulting services can be efficient
in many dynamic decision problems.

These results hold when the information provider's discount factor is equal to or less than the DM's discount factor.
However, when the information provider adopts a greater discount factor, the upfront fee is no longer optimal.
In this case, the optimal fee scheme involves charging a large lump sum fee and delaying long enough the payment time.

We also study the implications of information quality on the optimal stopping/waiting strategies.
We show that in terms of optimal strategies, having access to high quality information is similar to having a greater discount factor:
both have the effect of increasing the optimal stopping thresholds and making the DM more prone to wait.
Nevertheless, in terms of expected stopping time and stopping time distribution, better information and a greater discount factor have different effects.
Under a greater discount factor, the induced stopping time distribution first-order stochastically dominates the stopping time distribution under
a lower one, hence the expected stopping time associated to a greater discount factor is longer.
In contrast, having access to better information may either shorten or lengthen the expected stopping time.
The stopping time distribution induced by a more informative information structure neither first-order nor second-order stochastically dominates
the one under an inferior information structure.

\bibliographystyle{newapa}
\bibliography{Literature_Patience_2021}

\begin{thebibliography}{}

\bibitem[\protect\citeauthoryear{Arrow \& Fisher}{Arrow \&
  Fisher}{1974}]{Arrow1974}
Arrow, K.~J. \& Fisher, A.~C. (1974).
\newblock Environmental preservation, uncertainty, and irreversibility.
\newblock {\em The Quarterly Journal of Economics}, {\em 88\/}(2), 312--319.

\bibitem[\protect\citeauthoryear{Azrieli \& Lehrer}{Azrieli \&
  Lehrer}{2008}]{Azrieli2008}
Azrieli, Y. \& Lehrer, E. (2008).
\newblock The value of a stochastic information structure.
\newblock {\em Games and Economic Behavior}, {\em 63\/}(2), 679--693.

\bibitem[\protect\citeauthoryear{Babaioff, Kleinberg \& Paes~Leme}{Babaioff
  et~al.}{2012}]{babaioff2012optimal}
Babaioff, M., Kleinberg, R., \& Paes~Leme, R. (2012).
\newblock Optimal mechanisms for selling information.
\newblock In {\em Proceedings of the 13th ACM Conference on Electronic
  Commerce}, (pp.\ 92--109).

\bibitem[\protect\citeauthoryear{Bergemann \& Bonatti}{Bergemann \&
  Bonatti}{2015}]{Bergemann2015}
Bergemann, D. \& Bonatti, A. (2015).
\newblock Selling cookies.
\newblock {\em American Economic Journal: Microeconomics}, {\em 7\/}(3),
  259--294.

\bibitem[\protect\citeauthoryear{Bergemann, Bonatti \& Smolin}{Bergemann
  et~al.}{2018}]{Bergemann2018}
Bergemann, D., Bonatti, A., \& Smolin, A. (2018).
\newblock The design and price of information.
\newblock {\em American Economic Review}, {\em 108\/}(1), 1--48.

\bibitem[\protect\citeauthoryear{Blackwell}{Blackwell}{1953}]{Blackwell1953}
Blackwell, D. (1953).
\newblock Equivalent comparisons of experiments.
\newblock {\em The Annals of Mathematical Statistics}, {\em 24\/}(2), 265--272.

\bibitem[\protect\citeauthoryear{Blackwell}{Blackwell}{1965}]{Blackwell1965}
Blackwell, D. (1965).
\newblock Discounted dynamic programming.
\newblock {\em The Annals of Mathematical Statistics}, {\em 36\/}(1), 226--235.

\bibitem[\protect\citeauthoryear{Bloedel \& Zhong}{Bloedel \&
  Zhong}{2020}]{Bloedel2020}
Bloedel, A.~W. \& Zhong, W. (2020).
\newblock The cost of optimally acquired information.
\newblock {\em Working Paper, Stanford University}.

\bibitem[\protect\citeauthoryear{Cabrales, Gossner \& Serrano}{Cabrales
  et~al.}{2013}]{Cabrales2013}
Cabrales, A., Gossner, O., \& Serrano, R. (2013).
\newblock Entropy and the value of information for investors.
\newblock {\em American Economic Review}, {\em 103\/}(1), 360--377.

\bibitem[\protect\citeauthoryear{Che \& Mierendorff}{Che \&
  Mierendorff}{2019}]{Che2019}
Che, Y.-K. \& Mierendorff, K. (2019).
\newblock Optimal dynamic allocation of attention.
\newblock {\em American Economic Review}, {\em 109\/}(8), 2993--3029.

\bibitem[\protect\citeauthoryear{Chetty}{Chetty}{2007}]{Chetty2007}
Chetty, R. (2007).
\newblock Interest rates, irreversibility, and backward-bending investment.
\newblock {\em The Review of Economic Studies}, {\em 74\/}(1), 67--91.

\bibitem[\protect\citeauthoryear{Cukierman}{Cukierman}{1980}]{Cukierman1980}
Cukierman, A. (1980).
\newblock The effects of uncertainty on investment under risk neutrality with
  endogenous information.
\newblock {\em Journal of Political Economy}, {\em 88\/}(3), 462--475.

\bibitem[\protect\citeauthoryear{De~Lara \& Gossner}{De~Lara \&
  Gossner}{2017}]{DeLara2017}
De~Lara, M. \& Gossner, O. (2017).
\newblock An instrumental approach to the value of information.
\newblock {\em Working Paper}.

\bibitem[\protect\citeauthoryear{Demers}{Demers}{1991}]{Demers1991}
Demers, M. (1991).
\newblock Investment under uncertainty, irreversibility and the arrival of
  information over time.
\newblock {\em The Review of Economic Studies}, {\em 58\/}(2), 333--350.

\bibitem[\protect\citeauthoryear{Frankel \& Kamenica}{Frankel \&
  Kamenica}{2019}]{Frankel2019}
Frankel, A. \& Kamenica, E. (2019).
\newblock Quantifying information and uncertainty.
\newblock {\em American Economic Review}, {\em 109\/}(10), 3650--3680.

\bibitem[\protect\citeauthoryear{Gilboa \& Lehrer}{Gilboa \&
  Lehrer}{1991}]{Gilboa1991}
Gilboa, I. \& Lehrer, E. (1991).
\newblock The value of information---an axiomatic approach.
\newblock {\em Journal of Mathematical Economics}, {\em 20\/}(5), 443--459.

\bibitem[\protect\citeauthoryear{Kohlberg \& Neyman}{Kohlberg \&
  Neyman}{1999}]{Kohlberg1999}
Kohlberg, E. \& Neyman, A. (1999).
\newblock A strong law of large numbers for nonexpansive vector-valued
  stochastic processes.
\newblock {\em Israel Journal of Mathematics}, {\em 111\/}(1), 93--108.

\bibitem[\protect\citeauthoryear{Lehmann}{Lehmann}{1988}]{Lehmann1988}
Lehmann, E.~L. (1988).
\newblock {Comparing location experiments}.
\newblock {\em The Annals of Statistics}, {\em 16\/}(2), 521--533.

\bibitem[\protect\citeauthoryear{Lehrer \& Pauzner}{Lehrer \&
  Pauzner}{1999}]{Lehrer1999}
Lehrer, E. \& Pauzner, A. (1999).
\newblock Repeated games with differential time preferences.
\newblock {\em Econometrica}, {\em 67\/}(2), 393--412.

\bibitem[\protect\citeauthoryear{Liang, Mu \& Syrgkanis}{Liang
  et~al.}{2019}]{Liang2019}
Liang, A., Mu, X., \& Syrgkanis, V. (2019).
\newblock Optimal and myopic information acquisition.
\newblock {\em Working Paper}.

\bibitem[\protect\citeauthoryear{Mayskaya}{Mayskaya}{2019}]{Mayskaya2019}
Mayskaya, T. (2019).
\newblock Dynamic choice of information sources.
\newblock {\em Working Paper}.

\bibitem[\protect\citeauthoryear{Morris \& Strack}{Morris \&
  Strack}{2019}]{Morris2019}
Morris, S. \& Strack, P. (2019).
\newblock The {W}ald problem and the relation of sequential sampling and
  ex-ante information costs.
\newblock {\em Working Paper}.

\bibitem[\protect\citeauthoryear{Moscarini \& Smith}{Moscarini \&
  Smith}{2001}]{Moscarini2001}
Moscarini, G. \& Smith, L. (2001).
\newblock The optimal level of experimentation.
\newblock {\em Econometrica}, {\em 69\/}(6), 1629--1644.

\bibitem[\protect\citeauthoryear{Moscarini \& Smith}{Moscarini \&
  Smith}{2002}]{Moscarini2002}
Moscarini, G. \& Smith, L. (2002).
\newblock The law of large demand for information.
\newblock {\em Econometrica}, {\em 70\/}(6), 2351--2366.

\bibitem[\protect\citeauthoryear{Pindyck}{Pindyck}{1991}]{Pindyck1991}
Pindyck, R.~S. (1991).
\newblock Irreversibility, uncertainty, and investment.
\newblock {\em Journal of Economic Literature}, {\em 29\/}(3), 1110--1148.

\bibitem[\protect\citeauthoryear{Wald}{Wald}{1945}]{Wald1945}
Wald, A. (1945).
\newblock Sequential tests of statistical hypotheses.
\newblock {\em Annals of Mathematical Statistics}, {\em 16\/}(2), 117--186.

\bibitem[\protect\citeauthoryear{Wald}{Wald}{1947}]{Wald1947}
Wald, A. (1947).
\newblock Foundations of a general theory of sequential decision functions.
\newblock {\em Econometrica}, {\em 15\/}(4), 279--313.

\end{thebibliography}

\appendix
\section{Appendix}


\subsection{Proof of Proposition \ref{Prop_decreasing stopping sets}}
We prove by induction that $V^N_n(\mu)\geq V^N_{n+1}(\mu)$. For the last period, $V^N_N(\mu)= \max\{\pi(\mu),0\}\geq 0$.
Hence, $V^N_{N-1}(\mu)=\max\{\pi(\mu),\delta[\E(V^N_N(\mu(s))|\mu)-c]\}\geq V^N_N(\mu)$.
Now suppose $V^N_n(\mu)\geq V^N_{n+1}(\mu)$. It follows that $V^N_n(\mu)\geq V^N_{n+1}(\mu)$, and that
$\E(V^N_n(\mu(s))|\mu)\geq \E(V^N_{n+1}(\mu(s))|\mu)$.
Hence each component function of $V^N_{n-1}(\mu)$ is weakly greater than that of the $V^N_n(\mu)$,
therefore, $V^N_{n-1}(\mu)\geq V^N_n(\mu)$.

To show that the stopping sets satisfy $E^N_n\subseteq E^N_{n+1}$, note that, by definition,
$E^N_n = \{\mu\in\Delta\Theta|~ \pi(\mu)=V^N_n(\mu)\}$, $E^N_{n+1} = \{\mu\in\Delta\Theta|~ \pi(\mu)=V^N_{n+1}(\mu)\}$.
The desired result follows from the inequality $V^N_n(\mu)\geq V^N_{n+1}(\mu)$.


\subsection{Proof of Lemma \ref{Lemma_Decompose value function}}

Let $\{A^N_n(h_n;c)\},\  \{E^N_n(h_n;c)\}$ be the of optimal information acquisition and stopping sets corresponding to the fee function $c$.

In view of Eqs. (\ref{Value_Func_V_n}) and (\ref{Value_Func_V_0}),
\begin{eqnarray}
V^N_0(\mu_0;\varnothing;c) &=& \pi(\mu_0) \mathbbm{1}_{\left\{ \mu_0\in E^N_0(\varnothing;c) \right\}}  -  c(\varnothing) \mathbbm{1}_{ \left\{ \mu_0\in A^N_0(\varnothing;c) \right\} }    \notag \\
                           & & + ~ \delta \left[ \E\left( V^N_1(\mu_0(h_1);h_1;c)|\mu_0;\varnothing \right) - E\left( c(h_1) | \mu_0;\varnothing \right)  \right]  \mathbbm{1}_{ \left\{ \mu_0\in A^N_0(\varnothing;c) \right\} }     \notag \\
                           &=& \pi(\mu_0) \mathbbm{1}_{\left\{ \mu_0\in E^N_0(\varnothing;c) \right\}} + \delta \E\left( V^N_1(\mu_0(h_1);h_1;c)|\mu_0;\varnothing \right) \mathbbm{1}_{\left\{ \mu_0\in A^N_0(\varnothing;c) \right\}}  \notag \\
                           & & -  ~ c(\varnothing) \mathbbm{1}_{ \left\{ \mu_0\in A^N_0(\varnothing;c) \right\} } - \delta \sum_{h_1\in H_1} c(h_1) \P(h_1|\mu_0) \mathbbm{1}_{\left\{ \mu_0\in A^N_0(\varnothing;c) \right\}}.   \label{Eq_Decomposition_1}
\end{eqnarray}
Consider the term $\delta \E\left( V^N_1(\mu_0(h_1);h_1;c)|\mu_0;\varnothing \right) \mathbbm{1}_{\left\{ \mu_0\in A^N_0(\varnothing;c) \right\}}$
in Eq.\ (\ref{Eq_Decomposition_1}). Note that
\begin{eqnarray}
 & & \E\left( V^N_1(\mu_0(h_1);h_1;c)|\mu_0;\varnothing \right) \mathbbm{1}_{\left\{ \mu_0\in A^N_0(\varnothing;c) \right\}}  \notag \\
 &=& \sum_{h_1\in H_1} \pi(\mu_0(h_1)) \P(h_1|\mu_0) \mathbbm{1}_{\left\{\mu_0(h_1)\in E^N_1(h_1;c), ~ \mu_0\in A^N_0(\varnothing;c) \right\}}  \notag \\
 & & + \delta \sum_{h_1\in H_1} \left[ \E(V^N_2(\mu_0(h_2);h_2;c)| \mu_0,h_1) - \E(c(h_2)|\mu_0;h_1)  \right] \P(h_1|\mu_0) \mathbbm{1}_{\left\{\mu_0(h_1)\in A^N_1(h_1;c), ~ \mu_0\in A^N_0(\varnothing;c) \right\}}  \notag  \\
 &=& \sum_{h_1\in H_1} \pi(\mu_0(h_1))  \P(h_1|\mu_0) \mathbbm{1}_{\left\{\mu_0(h_1)\in E^N_1(h_1;c), ~ \mu_0\in A^N_0(\varnothing;c) \right\}}  \notag \\
 & & + ~ \delta \sum_{h_1\in H_1} \E(V^N_2(\mu_0(h_2);h_2;c)| \mu_0,h_1) \P(h_1|\mu_0) \mathbbm{1}_{\left\{\mu_0(h_1)\in A^N_1(h_1;c), ~ \mu_0\in A^N_0(\varnothing;c) \right\}} \notag \\
 & & - ~ \delta \sum_{h_2\in H_2} c(h_2) \P(h_2|\mu_0) \mathbbm{1}_{\left\{\mu_0(h_1)\in A^N_1(h_1;c), ~ \mu_0\in A^N_0(\varnothing;c) \right\}}. \notag
\end{eqnarray}
Plugging this expression in Eq.\ (\ref{Eq_Decomposition_1}), we obtain
\begin{eqnarray}
V^N_0(\mu_0;\varnothing;c) &=& \pi(\mu_0) \mathbbm{1}_{\left\{ \mu_0\in E^N_0(\varnothing;c) \right\}} + \delta \sum_{h_1\in H_1} \pi(\mu_0(h_1)) \P(h_1|\mu_0) \mathbbm{1}_{\left\{\mu_0(h_1)\in E^N_1(h_1;c), ~ \mu_0\in A^N_0(\varnothing;c) \right\}} \notag \\
                           & & + ~ \delta^2 \sum_{h_1\in H_1} \E(V^N_2(\mu_0(h_2);h_2;c)| \mu_0,h_1) \P(h_1|\mu_0) \mathbbm{1}_{\left\{\mu_0(h_1)\in A^N_1(h_1;c), ~ \mu_0\in A^N_0(\varnothing;c) \right\}}   \notag \\
                           & & - ~ c(\varnothing) \mathbbm{1}_{ \left\{ \mu_0\in A^N_0(\varnothing;c) \right\} } - \delta \sum_{h_1\in H_1} c(h_1) \P(h_1|\mu_0) \mathbbm{1}_{\left\{ \mu_0\in A^N_0(\varnothing;c) \right\}}   \notag \\
                           & & - ~ \delta^2 \sum_{h_2\in H_2} c(h_2) \P(h_2|\mu_0) \mathbbm{1}_{\left\{\mu_0(h_1)\in A^N_1(h_1;c), ~ \mu_0\in A^N_0(\varnothing;c) \right\}}.   \label{Eq_Decomposition_2}
\end{eqnarray}
In Eq.\ (\ref{Eq_Decomposition_2}), the first line consists of the discounted expected entry payoffs when entry occurs in period 0 and period $1$, respectively,
following the optimal strategy when the fee function is $c$ (note that exits yield payoff 0);
the second line is the discounted expected continuation value conditional on no entry in the first two periods;
the last two lines correspond to the discounted expected costs incurred in the first two periods following the optimal information acquisition strategy
characterized by $\{A^N_n(h_n;c)\}$.

Continuing inductively in this fashion, the collection of all terms that involve the costs is exactly $-\rho(\mu_0;c)$  (see Eq.\ (\ref{Total expected cos under c})).
The sum of the remaining terms is the discounted expected gross payoff when facing the cost function $\pmb{0}$,
using the strategy characterized by $\{A^N_n(h_n;c)\}$ and  $\{B^N_n(h_n;c)\}$,
or $\hat{V}^N_0(\mu_0;\varnothing;\pmb{0})$.P


\subsection{Proof of Proposition \ref{Prop_V_n_convex}}
Continuity is clear. To establish convexity,
we first prove that for any period $n$,
the convexity of $V_{n+1}^N(\mu;h_{n+1})$ for each $h_{n+1}$ would imply that
$\E(V_{n+1}^N(\mu(s);h_{n+1})|\mu;h_n)$ is convex in $\mu$ for each $h_n$.
We then establish by induction that for each $n$ and $h_n\in H_n$, $V^N_n(\mu;h_n)$ is indeed convex.

Take any two different beliefs $\mu,\ \mu'\in\Delta\Theta$, and any $\lambda\in(0,1)$. Set $\tilde{\mu}:=\lambda \mu + (1-\lambda)\mu'$.
Let $f_{\theta}(s)$ be the probability of receiving the signal $s$ under state $\theta$. Define
$\alpha_s(\tilde{\mu}):=\sum_{\theta\in\Theta} \tilde{\mu}_\theta f_{\theta}(s)$,
which is the probability of getting signal $s$ under belief $\tilde{\mu}$. Clearly
$$\alpha_s(\tilde{\mu}) = \sum_{\theta\in\Theta}\left(\lambda\mu_{\theta} + (1-\lambda)\mu_{\theta}'\right) f_{\theta}(s) = \lambda \alpha_s(\mu) + (1-\lambda)\alpha_s(\mu').$$
Given a belief $\mu$, let $\mu(s)\in \Delta\Theta$ be the posterior belief after observing the signal $s$.
With these notations, we have
\begin{eqnarray}
& & \lambda \E\left(V_{n+1}^N(\mu(s);h_{n+1})|\mu;h_n \right) + (1-\lambda) \E\left( V_{n+1}^N(\mu'(s);h_{n+1})|\mu';h_n\right)   \notag \\
&=& \lambda \sum_s \alpha_s(\mu) V_{n+1}^N(\mu(s);h_{n+1}) + (1-\lambda)\sum_s \alpha_s(\mu') V_{n+1}^N(\mu'(s);h_{n+1})            \notag \\
&=& \sum_s \alpha_s(\tilde{\mu}) \left( \frac{\lambda\alpha_s(\mu)}{\alpha_s(\tilde{\mu})} V_{n+1}^N(\mu(s);h_{n+1})  +  \frac{(1-\lambda) \alpha_s(\mu')}{\alpha_s(\tilde{\mu})}V_{n+1}^N(\mu'(s);h_{n+1})\right) \notag \\  
&\geq&  \sum_s \alpha_s(\tilde{\mu}) \left( V^N_{n+1}\left( \frac{\lambda\alpha_s(\mu)}{\alpha_s(\tilde{\mu})} \mu(s) +  \frac{(1-\lambda)\alpha_s(\mu')}{\alpha_s(\tilde{\mu})} \mu'(s); h_{n+1}  \right)  \right),   \label{Prop_V_convex_equi_1}
\end{eqnarray}
where the inequality is due to the assumption that for each history $h_{n+1}$, $V^N_{n+1}(\mu;h_{n+1})$ is convex in the belief $\mu$.
Notice that the posteriors $\mu(s),\ \mu'(s),\ \tilde{\mu}(s)$ are vectors of the following forms:
\begin{equation*}
  \mu(s)=\frac{\left( \mu_\theta f_\theta(s) \right)_{\theta\in\Theta}}{\alpha_s(\mu)}, \ \  \mu'(s)=\frac{ \left( \mu'_\theta f_\theta(s)\right)_{\theta\in\Theta}}{\alpha_s(\mu')},\ \  \tilde\mu(s)=\frac{\left( \tilde\mu_\theta f_\theta(s) \right)_{\theta\in\Theta}}{\alpha_s(\tilde\mu)}.
\end{equation*}
Therefore, in Eq. (\ref{Prop_V_convex_equi_1}),
\begin{eqnarray}
   \frac{\lambda\alpha_s(\mu)}{\alpha_s(\tilde{\mu})} \mu(s) +  \frac{(1-\lambda)\alpha_s(\mu')}{\alpha_s(\tilde{\mu})} \mu'(s) &=& \frac{\left( \lambda\mu_\theta f_\theta(s) + (1-\lambda)\mu'_\theta f_\theta(s) \right)_{\theta\in\Theta}}{\alpha_s(\tilde{\mu})} \notag \\
                                                                                                                                &=& \frac{\left( \tilde\mu_{\theta} f_\theta(s) \right)_{\theta\in\Theta}}{\alpha_s(\tilde\mu)} \notag \\
                                                                                                                                &=& \tilde\mu(s), \notag
\end{eqnarray}
and so Eq. (\ref{Prop_V_convex_equi_1}) reduces to
\begin{eqnarray*}
&    & \lambda \E\left(V_{n+1}^N(\mu(s);h_{n+1})|\mu;h_n \right) + (1-\lambda) \E\left( V_{n+1}^N(\mu'(s);h_{n+1})|\mu';h_n\right)  \\
&\geq& \sum_s \alpha_s(\tilde\mu) V^N_{n+1}(\tilde\mu(s);h_{n+1})  \\
& =  & \E\left( V^N_{n+1}(\tilde\mu(s);h_{n+1})|\mu;h_n \right).
\end{eqnarray*}
This establishes that $\E(V_{n+1}^N(\mu(s);h_{n+1})|\mu;h_n)$ is convex in $\mu$ whenever $V_{n+1}^N(\mu;h_{n+1})$ is convex in $\mu$.

Now we prove that $V_{n}^N(\mu;h_{n})$ is indeed convex in $\mu$ for each $n$ and $h_n\in H_n$. Start with the last period.
Since the expected entry payoff $\pi(\mu):=\max_{a\in A}\sum_{\theta\in\Theta} \mu_\theta u(\theta,a)$ is convex in $\mu$, it follows that
$V_{N}^N(\mu;h_N)=\max\{\pi(\mu),0\}$ is convex in $\mu$, hence by what we just proved, $\E(V^N_{N}(\mu(s);h_N)|\mu;h_{N-1})$
is convex in $\mu$ for any $h_{N-1}\in H_{N-1}$.
Therefore, the value function
$$V^N_{N-1}(\mu;h_{N-1})=\max\left\{\pi(\mu), \delta \left[ \E(V^N_N(\mu(s);h_{N})|\mu;h_{N-1}) - \E(c(h_N)|\mu; h_{N-1}) \right] \right\}$$
is convex, since each component function in the $\max$ operator is convex in $\mu$ for any $h_{N-1}\in H_{N-1}$ (in particular,
$\E(c(h_N)|\mu; h_{N-1})=\sum_s \sum_{\theta\in\Theta} \mu_\theta f_{\theta}(s)c(h_{N}|h_{N-1})$ is linear in $\mu$).
Therefore, $\E(V^N_{N-1}(\mu(s);h_{N-1})|\mu;h_{N-2})$ is convex in $\mu$ for each $h_{N-2}\in H_{N-2}$.
By induction, all value functions $V^N_n(\mu;h_n)$ are convex in $\mu$.


\subsection{Proof of Proposition \ref{Theorem_compare entry sets}} \label{Appendix_proof of theorem}
The proof for part 1 follows immediately from the fact that any signal from the information structure $S$
can be garbled and regarded as a $T$-signal (Definition \ref{Def_More_precise_info_structure}).
Then acting on this signal with the optimal strategy of $T$ guarantees the value under $T$.
Alternatively, one can show by applying Blackwell's equivalence theorem.

To show that $E^N_n(S)\subseteq E^N_n(T)$ and $A^N_n(T)\subseteq A^N_n(S)$, note that, by definition,
$E^N_n(S)= \{\mu\in\Delta\Theta| ~ V^N_n(\mu)=\pi(\mu)\}$, $E^N_n(T)= \{\mu\in\Delta\Theta|~ U^N_n(\mu)=\pi(\mu)\}$,
where $U^N_n(\mu)=\max\{\pi(\mu),\delta\E(U^N_{n+1}(\mu(t)) |\mu) \}$, $V^N_n(\mu)=\max\{\pi(\mu),\delta\E(V^N_{n+1}(\mu(s)) |\mu) \}$.
It follows from part 1 of the proposition that when $V^N_n(\mu)=\pi(\mu)$,
\begin{equation*}
  \pi(\mu) \geq \delta \E(V^N_{n+1}(\mu(s)) |\mu) \geq \delta \E(U^N_{n+1}(\mu(t)) |\mu),
\end{equation*}
hence $U^N_n(\mu) =\pi(\mu)$. Therefore, $E^N_n(S)\subseteq E^N_n(T)$. Because under the upfront fee scheme, the information acquisition set is
the complement of the stopping set, we conclude that $A^N_n(T)\subseteq A^N_n(S)$.


\subsection{Proof of Proposition \ref{Prop_Waiting set and delta}}
In the proof, we consider the case when the DM faces a general history-dependent fee scheme $c(h_n)$.
Let $V^N_n(\mu;h_n;\delta)$ and $V^N_n(\mu;h_n;\hat\delta)$ be the value functions under $\delta$ and $\hat\delta$, respectively, where $\delta >\hat\delta$.
By induction we show that for any $ n$  and any $h_n$, $V^N_n(\mu;h_n;\delta)\geq V^N_n(\mu;h_n;\hat\delta)$.
For the last period, $V^N_N(\mu;h_n;\delta)=V^N_N(\mu;h_n;\hat\delta)=\pi(\mu)$.
Assume that the claim holds for $n+1$.
We show that each component function of $V^N_n(\mu;h_n;\delta)$ is greater than the corresponding one of $V^N_n(\mu;h_n;\hat{\delta})$.
Clearly, $\delta V^N_{n+1}(\mu;h_{n+1};\delta) \geq \hat{\delta} V^N_{n+1}(\mu;h_{n+1}; \hat{\delta})$.
For the continuation value following a history $h_n$ when the DM chooses to acquire information,
\begin{eqnarray*}
\E\left( V_{n+1}^N(\mu(s);h_{n+1};\delta)|\mu;h_n\right) & =  & \sum_{s} \alpha_s(\mu) V^N_{n+1}(\mu(s);h_{n+1};\delta) \\
                                                         &\geq& \sum_{s} \alpha_s(\mu) V^N_{n+1}(\mu(s);h_{n+1};\hat{\delta}) \\
                                                         & =  & \E( V_{n+1}^N(\mu(s);h_{n+1};\hat{\delta})|\mu;h_n).
\end{eqnarray*}
Consequently, $V^N_n(\mu;h_n;\delta)\geq V^N_n(\mu;h_n;\hat\delta)$.

The optimal stopping set under $\delta$ is characterized by $E^N_n(h_n;\delta)=\{\mu\in\Delta\Theta|~ \pi(\mu)=V^N_n(\mu;h_n;\delta)\}$.
Since $V^N_n(\mu;h_n;\delta)\geq V^N_n(\mu;h_n;\hat{\delta})$, it follows that $E^N_n(h_n;\delta)\subseteq E^N_n(h_n;\hat{\delta})$ for any history $h_n$.
As a result, $A^N_n(h_n;\hat{\delta})\cup W^N_n(h_n;\hat{\delta}) \subseteq A^N_n(h_n;\delta)\cup W^N_n(h_n;\delta)$.
When we restrict attention to the upfront fee scheme, the waiting sets $W^N_n(\cdot)$ are empty,
so we have the result stated in Proposition \ref{Prop_Waiting set and delta}.

Now we show that the stopping time distribution induced by $\delta$ first-order stochastically dominates that induced by $\hat{\delta}$.
Let $H_n(\delta)$ and $H_n(\hat{\delta})$ be the sets of length-$n$ histories that lead to waiting (with or without acquiring information) up to period $n$
under $\delta$ and $\hat{\delta}$, respectively. Since the waiting set in every period is larger under $\delta$ than under $\hat{\delta}$,
it follows that $H_n(\hat{\delta})\subseteq H_n(\delta)$. Consequently, $\P(H_n(\delta))\geq \P(H_n(\hat{\delta}))$.
It says that the probability that stopping does not occur up to period every $n$ is greater under $\delta$.
This completes the proof that the stopping time distribution induced by $\delta$ first-order stochastically dominates that induced by $\hat{\delta}$.


\subsection{Proof of Lemma \ref{Lemma_Infitnite horizon}}
We show that the mapping $H$ defined by
\begin{equation*}
H(V(\mu)) = \max\left\{\pi(\mu),\delta \E(V(\mu(s))|\mu)\right\}
\end{equation*}
is a contraction mapping.
To this end, we verify that the mapping $H(\cdot)$ satisfies the following
two Blackwell sufficient conditions for contraction mapping (Theorem 5, \citet{Blackwell1965}):
\begin{enumerate}
  \item \emph{Monotonicity}. For any two functions $V(\cdot),\ V'(\cdot)$, if $V(\mu)\geq V'(\mu)$, $\forall \mu\in\Delta\Theta$,
        then $H(V(\mu))\geq H(V'(\mu))$.
  \item \emph{Discounting}. For any non-negative constant $a$, $H((V+a)(\mu)) \leq H(V(\mu)) + \delta a$, where $(V+a)(\mu)$ is the function defined
        by $(V+a)(\mu):=V(\mu)+a$.
\end{enumerate}

Take any two functions $V(\mu)$ and $V'(\mu)$ with $V(\mu)\geq V'(\mu)$, $\forall \mu$.
Then $\E(V(\mu(s))|\mu)-\E(V'(\mu(s))|\mu)\geq 0$, hence $H(V(\mu))\geq H(V'(\mu))$, $\forall \mu$, which establishes the monotonicity of the mapping $H(\cdot)$.

Now let us show that the mapping $H(\cdot)$ also satisfies discounting.
By definition, $H((V+a)(\mu)) = \max\left\{\pi(\mu), \delta \E(V(\mu(s))|\mu) + \delta a\right\}$.
There are two possible cases: either $\delta \E(V(\mu(s))|\mu)\geq \pi(\mu)$, or
$\delta \E(V(\mu(s))|\mu) < \pi(\mu)$.
In the former case, $H((V+a)(\mu)) = \delta \E(V(\mu(s))|\mu)+\delta a = H(V(\mu)) + \delta a$, which satisfies discounting.
In the latter case, we have $H(V(\mu))= \pi(\mu)$.
It follows that $H(V(\mu))+\delta a = \pi(\mu) + \delta a > \delta \E(V(\mu(s))|\mu) + \delta a$,
hence $H(V(\mu))+\delta a > \max\left\{\pi(\mu), \delta \E(V(\mu(s))|\mu)+\delta a\right\} =H((V+a)(\mu))$.

We conclude that the mapping $H(\cdot)$ is indeed a contraction mapping. Therefore, there exists a unique fixed point
$V(\cdot)$ that satisfies Eq.\ (\ref{V_Bellman_Equ_Infiinte}), which is the limit of the sequence of value functions $\{V_0^N(\mu)\}_{N=0}^\infty$.

\end{document}